\documentclass[aps,prl,twocolumn,notitlepage,superscriptaddress]{revtex4-1}
\usepackage[colorlinks=true, citecolor=magenta, linkcolor=blue, urlcolor=blue]{hyperref}
\usepackage{graphicx}
\usepackage{amsmath}
\usepackage{amssymb}
\usepackage{amsfonts}
\usepackage{hyperref}
\usepackage{mathtools}
\usepackage{xcolor}
\usepackage{verbatim}

\usepackage{tikz}
\usetikzlibrary {shapes.geometric}

\usepackage{color}
\usepackage{comment}

\newcommand{\eps}[0]{\epsilon}
\newcommand{\ket}[1]{\left| #1 \right. \rangle}
\newcommand{\bra}[1]{\langle \left. #1 \right|}
\newcommand{\tr}[0]{\mathrm{Tr}}
\newcommand{\cdag}{c^{\dagger}}

\begin{document}
\title{Super-Poissonian noise from quasiparticle poisoning in electron transport through a pair of Majorana bound states}
\author{Florinda Vi\~nas Bostr\"om}
\affiliation{Institut f\"ur Mathematische Physik, Technische Universit\"at Braunschweig, D-38106 Braunschweig, Germany}
\affiliation{Division of Solid State Physics and NanoLund, Lund University, Box 118, S-221 00 Lund, Sweden}
\author{Patrik Recher}
\affiliation{Institut f\"ur Mathematische Physik, Technische Universit\"at Braunschweig, D-38106 Braunschweig, Germany}
\affiliation{Laboratory of Emerging Nanometrology, D-38106 Braunschweig, Germany}
\date{\today} 

\begin{abstract}
Topological qubits based on non-abelian Majorana bound states (MBSs) are protected by parity which is challenged by quasiparticle poisoning (QPP).
In this work, we show how QPP affects transport through a pair of coupled MBSs weakly connected to two current leads, using an open system approach and full counting statistics.
We find that the correct low-energy physics requires to include next to leading order tunneling events in the coupling to the leads. In particular, our results show that QPP causes super-Poissonian local shot noise with a Fano factor that diverges with decreasing bias voltage, while the current and the non-local (cross-)noise are only little affected. We explain that these features are a direct consequence of the nature of MBSs being their own antiparticle making noise measurements a viable tool to search for MBSs in the presence of QPP.
\end{abstract}

\maketitle

Majorana bound states (MBSs) are predicted to be zero-energy excitations at the boundary of topological superconductors \cite{Read2000, Kitaev2001, Fu2008, Oreg2010, Lutchyn2010, Leijnse2012, Beenakker2013b, Aguado2017, Prada2020a} with the defining property that its creation operator $\gamma^{\dagger}$ is identical to its annihilation operator $\gamma$, but their existence has not been conclusively confirmed by experiments yet, even though they should be topologically protected, hence robust to disorder. 

These non-abelian quasiparticles have been advertised as building blocks for topological quantum computing \cite{Nayak2008, Kitaev2003, Fu2008, Aasen2016, Hoffman2016, Landau2016, Plugge2017}. The superior power of such qubits lies in the non-local encoding of quantum information. These qubits, however, are prone to parity changes, i.e., number of particles in the superconductor being odd or even. Parity switching events are therefore detrimental for topological quantum computing. An exception is their controlled use in braiding processes of MBSs \cite{Park2020}. However, uncontrolled (i.e. random)  parity changes appear naturally in open quantum systems and are also referred to as quasiparticle poisoning (QPP) \cite{Goldstein2011,Budich2012,Rainis2012,Plugge2016,Albrecht2017, Schulenborg2023}.
The QPP mechanism is relevant not only for systems hosting MBSs, but also for superconducting (non-topological) qubits \cite{Aumentado2004,Lutchyn2005,Catelani2011,Liu2024}, and for quantum dot based topological superconductor devices \cite{Tsintzis2024}.

In this letter, we consider the transport properties of a pair of MBSs comprising a non-local fermion state as implemented, e.g., in a Majorana wire and attached to two leads, see Fig.~1(a). We apply an open systems approach for the calculation of the cumulant generating function (CGF) leading to average current and noise properties (and in principle to higher-order cumulants) in the presence of dissipation. In particular, we study the effects of QPP.

\begin{figure}[t]
        \begin{flushleft}
\large{(a)}
\end{flushleft} 
 \vspace{-0.8cm}
   \includegraphics[width=0.42\textwidth,  trim=7cm 24cm 6.9cm 1cm, clip]{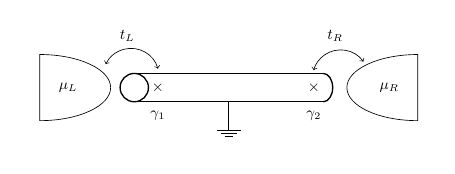} 
   \vspace{-0.5cm}
   \centering
      \begin{flushleft}
\large{(b)}
\end{flushleft} 
 \vspace{-1.4cm}
  \includegraphics[width=0.42\textwidth,  trim=7cm 25cm 7cm 0cm, clip]{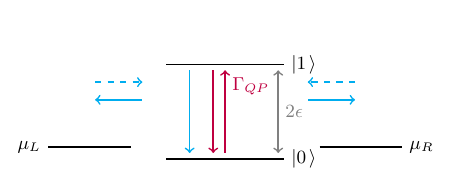} 
\caption{(a) Sketch of the transport setup.  (b) Energy diagram with possible ST at low chemical potentials $\mu_L$ and $\mu_R$. Deexciting the MBSs system (vertical blue arrow) can be associated with both tunneling out from the wire, via normal tunneling (blue horizontal arrows), or into the wire, via anomalous tunneling (blue horizontal vertical arrows). QPP (dark red arrows) switches the parity state with associated energy cost $2\eps$ without any tunneling processes. SOT processes are not shown here, as they do not change the parity. However, they can contribute to current and noise.}
\label{fig_setup}
\end{figure}

Transport through such and related setups has been extensively studied in the absence of QPP \cite{Bolech2007, Nilsson2008, Prada2012, Wu2012, Huetzen2012,Rainis2013, Zocher2013,Weithofer2014,vanHeck2016,Zazunov2016,Liu2017,Schuray2017,Prada2017a, Clarke2017, Dourado2024, Kleinherbers2023, Liu2015a,Liu2015b,Smirnov2022,Smirnov2024} theoretically and experimentally \cite{Mourik2012, Deng2012, Lee2013, Deng2016, Aghaee2023}. In the low voltage bias regime, below the hybridization energy of the two MBSs, and at weak coupling, the cross noise properties have been shown to be perfectly correlated, hinting at the splitting of a Cooper pair via two MBSs as the dominant transport process \cite{Nilsson2008, Weithofer2014}. The local Fano factor is therefore unity. This physics has been attributed to the coupling of the tunneling electrons to single MBSs. In this work, we show that in the same low bias regime, the inclusion of QPP can lead to a dramatic change of these results, in particular, to a diverging super-Poissonian local Fano factor for the shot noise. We attribute this physics to the possible non-equilibrium population of the higher non-local fermion state of the MBS pair driven by the QPP rate and of the fact that a MBS couples to electrons and holes with the same strength by virtue of $\gamma=\gamma^{\dagger}$.

All calculations are performed analytically, yielding expressions for the transport quantities that are discussed in the results sections. We also find that the low bias regime cannot be described by lowest order tunneling (sequential tunneling, ST) events alone. Second-order tunneling (SOT) events (Andreev tunneling and electron cotunneling) we treat by deriving an effective Hamiltonian.

\emph{Model Hamiltonian}---
A pair of coupled MBSs, each residing at an end of a proximitized quantum wire put to ground is tunnel-coupled to two normal metallic leads, see Fig.~\ref{fig_setup}(a). 

The Hamiltonian for the system is given by (we set $\hbar=1$)
$H = H_{\gamma} + H_B + H_T$,
where 
\begin{equation}
H_{\gamma} =  i \eps \gamma_1 \gamma_2 = -\eps(1-2c^{\dagger}c)
\end{equation} 
describes the two coupled MBSs with self-adjoint operators $\gamma_1$,$\gamma_2$ and hybridization energy $\epsilon$ and where we have used fermion creation (annihilation) operators $c^{\dagger}$ ($c$) to rewrite the Majorana operators $\gamma_1 = c^{\dagger} + c$ and $\gamma_2 = i(c^{\dagger} - c)$. The  leads are modeled by the bath Hamiltonian
\begin{equation}
H_B = iv_F \sum_{i=L,R}\int_{-\infty}^{\infty} dx_i \psi^{\dagger}_{i}(x_i) \partial_{x_i} \psi_{i}(x_i),
\end{equation} 
where $\psi_{i}(x)$ ($\psi_i^{\dagger}(x)$) annihilate (create) chiral electrons at position $x$ in lead $i=L,R$ moving with the Fermi velocity $v_F$. The MBSs are tunnel-coupled to leads by the tunneling Hamiltonian
\begin{equation}
\begin{split}
    H_T =& t_L \psi_L \gamma_1  + t_R \psi_R\gamma_2 + h.c. \\
    = & t_L\psi_L\cdag + i t_R\psi_R \cdag + t_L\psi_Lc-it_R\psi_R c + h.c.,
\end{split}
\end{equation}
where $t_{R(L)}$ are amplitudes for tunneling of an electron at $x=0$ from lead $R(L)$ to the corresponding MBS, see Fig. 1(a). The first two terms of the second row correspond to normal tunneling, and the second two terms of the second row correspond to anomalous tunneling. The normal tunneling conserves charge in the system, while the anomalous tunneling only conserves parity. In the Supplemental Material~\cite{supplemental}, we consider a more general tunneling Hamiltonian, allowing for non-local tunneling amplitudes. 

{\it Master equation and full counting statistics}---
To calculate transport properties and to include dissipative processes like tunneling to the leads and QPP events we employ a Master equation approach and full counting statistics. We are mainly interested in the low-bias regime $eV \ll \epsilon$ where the coupling of electrons and holes to MBSs leads to peculiar transport properties like enhanced crossed Andreev reflection in the absence of QPP \cite{Nilsson2008, Weithofer2014}. ST events from the ground state $|0\rangle$ of $H_\gamma$ with $c|0\rangle=0$ to the occupied (excited) state $|1\rangle$ with $c^{\dagger} c|1\rangle=|1\rangle$ and energy $\epsilon$ are strongly suppressed (for $eV, k_BT \ll \epsilon$). SOT events, on the other hand, preserve the parity $\langle c^{\dagger} c\rangle$ and are dominant at small energies, see Fig. 2(a). 
To set up the full counting statistics, we introduce counting fields $\chi = \{ \chi_i \}$, counting particles in lead $i = L, R$.
To include SOT processes in the Master equation, we derive an effective Hamiltonian \cite{Auerbach1994, supplemental} to second order in $H_T$, leading to
$H^{eff}=H+H^{SOT}$,
with
\begin{equation}
\begin{split}
      H^{SOT} =& \ket{0}\bra{0} \sum_{a,b=1}^4(-1)^{a+1}\alpha_a \alpha_b F_a \frac{1}{H_B-E_0+2\eps}F_b \\
+ & \ket{1}\bra{1} \sum_{a,b=1}^4(-1)^{b+1}\alpha_a \alpha_b F_a \frac{1}{H_B-E_0-2\eps}F_b,
\end{split}
\end{equation}
where $a$ and $b$ run over the four terms in the original tunneling Hamiltonian $H_T$, $ \alpha_{a,b}  \in \{ t_L, t_R, t_L^*,t_R^*\}$ and $F_{a,b} \in \{ \psi_L, \psi_L^{\dagger}, \psi_R, \psi_R^{\dagger} \}$ are the corresponding bath field operators, and $E_0$ is the energy of the state of the bath.

The counting field dependent density matrix of the total system $\rho_{\rm tot}(\chi,t)$ satisfies a generalized von Neumann equation ${\dot \rho_{\rm tot}(\chi,t)}= -i{\mathcal L}_{H^{eff}}\rho_{\rm tot}(\chi,t)$ with the Liouvillian superoperator  ${\mathcal L}_{H^{eff}} X = (H^{eff}(\chi/2)X-XH^{eff}(-\chi/2))$. Here, we have introduced $H^{eff}(\chi)=\exp(-i\sum_i\chi_iN_i)H^{eff}\exp(i\sum_i\chi_iN_i)=H_B+H_\gamma+H_T(\chi)+H^{SOT}(\chi)$ with $N_i$ the number operator in lead $i$.
Following standard steps \cite{Blum1996,Probst2022} and assuming $\max\{eV,k_BT\}, 2\epsilon \gg \Gamma_i$, we derive a Master equation for the reduced density matrix $\rho(\chi,t)={\rm Tr}_B\rho_{\rm tot}(\chi,t)$ to lowest order in $H_T(\chi)+H^{SOT}(\chi)$ 
\begin{equation}
\dot{\rho}(\chi,t) = -\mathcal{L}(\chi)\rho(\chi,t),
\label{eq_ME_plain}
\end{equation}
where ${\rm Tr}_B$ means taking the trace over the bath degrees of freedom.
We observe that the non-diagonal elements of $\rho$ decouple, so that we can vectorize its diagonal entries $\langle n|\rho(\chi,t)|n\rangle\equiv \rho_{nn}(\chi,t)$ and the corresponding Liouvillian matrix $\mathcal{L}$ can be 
decomposed as
$\mathcal{L} = \mathcal{L}_{ST} + \mathcal{L}_{SOT}$,
with (in the basis $\left(\rho_{00}(\chi,t),\rho_{11}(\chi,t)\right)^{T}$)
\begin{equation}
\mathcal{L}_{ST} =   \begin{pmatrix}
   W_{10}( 0)  & -W_{01}(\chi) \\
     -W_{10}(\chi) &    W_{01}( 0)
    \end{pmatrix} 
\end{equation} accounting for the ST processes, and 
\begin{equation}
    \mathcal{L}_{SOT} = \begin{pmatrix}
        -W_{00}(\chi) & 0 \\
        0 & -W_{11}(\chi)
    \end{pmatrix}
\end{equation} for the SOT processes. 

\begin{figure}
    \centering
        \begin{flushleft}
\large{(a)}
\end{flushleft} 
\vspace{-1.2cm}
\includegraphics[width=0.42\textwidth,  trim=0.65cm 10cm 2.2cm 10cm, clip]{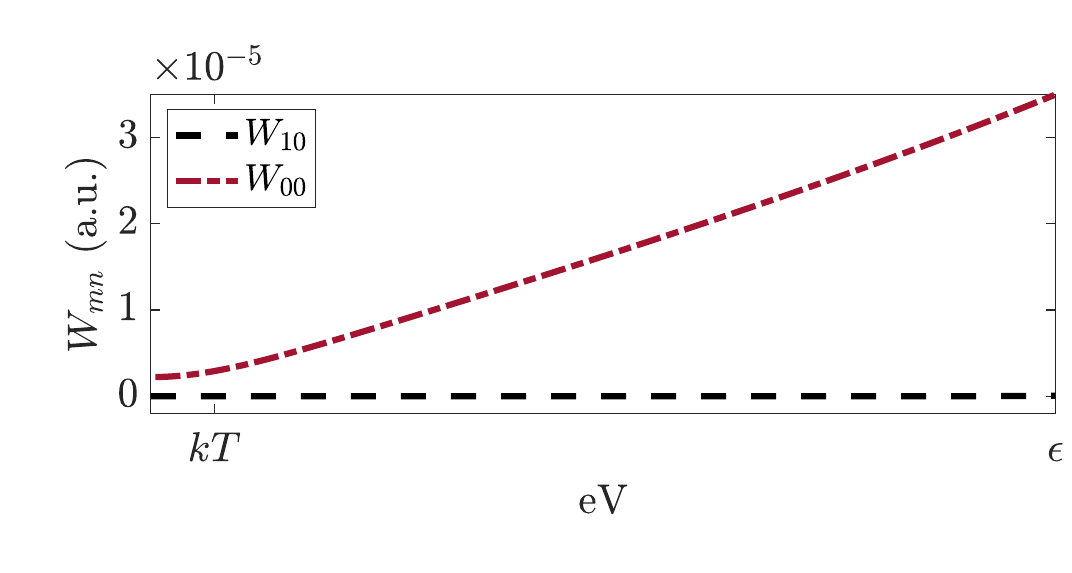}
\vspace{-1.2cm}
  \begin{flushleft}
 \large{(b)}
\end{flushleft} 
\vspace{-1.0cm}
\includegraphics[width=0.42\textwidth,  trim=0.65cm 9.5cm 2.2cm 10cm, clip]{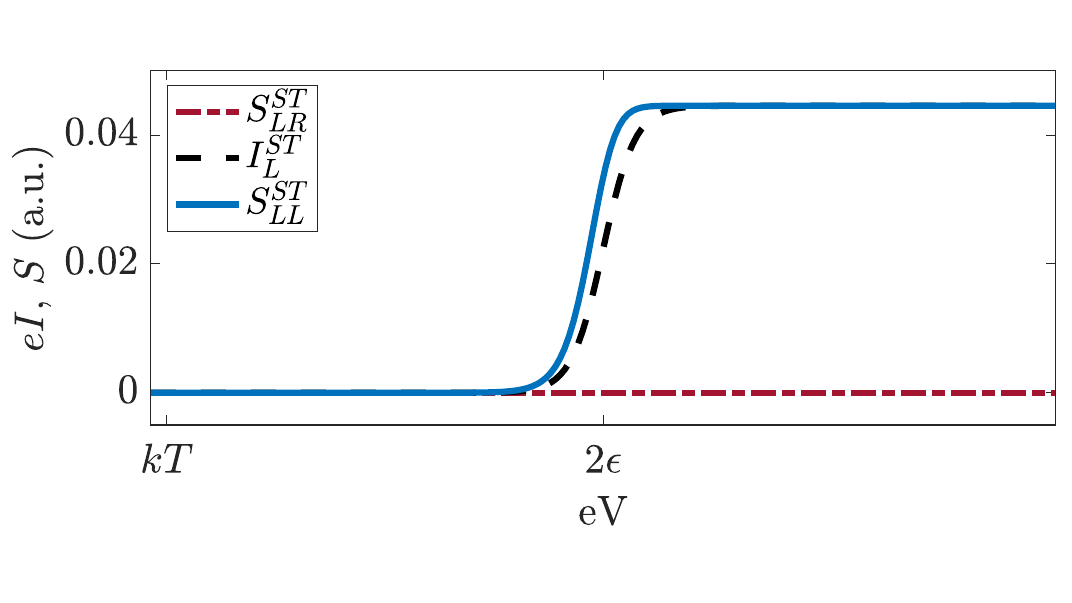}
\vspace{-1.0cm}
    \begin{flushleft}
\large{(c)}
\end{flushleft} 
 \vspace{-1cm}
     \includegraphics[width=0.42\textwidth,  trim=0.8cm 8.4cm 2.1cm 8.0cm, clip]{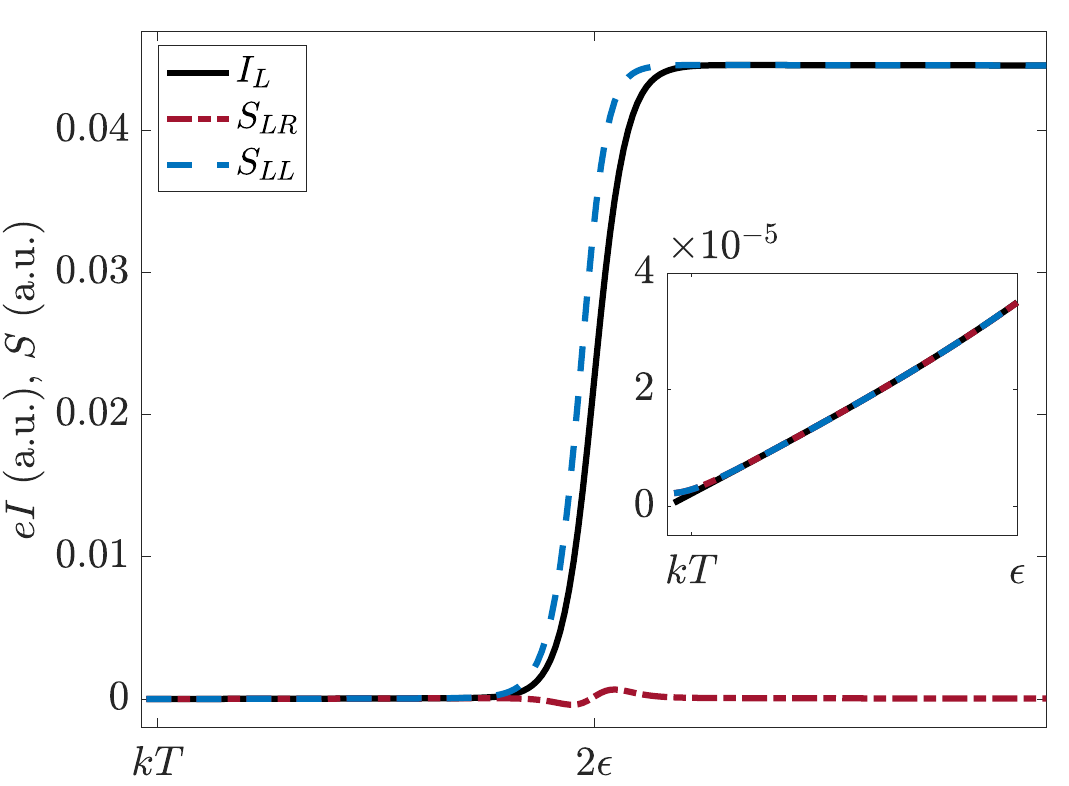} 
        \begin{flushleft}
\large{(d)}
\end{flushleft} 
 \vspace{-0.9cm}\includegraphics[width=0.42\textwidth,  trim=0.9cm 7.6cm 2.2cm 8.5cm, clip]{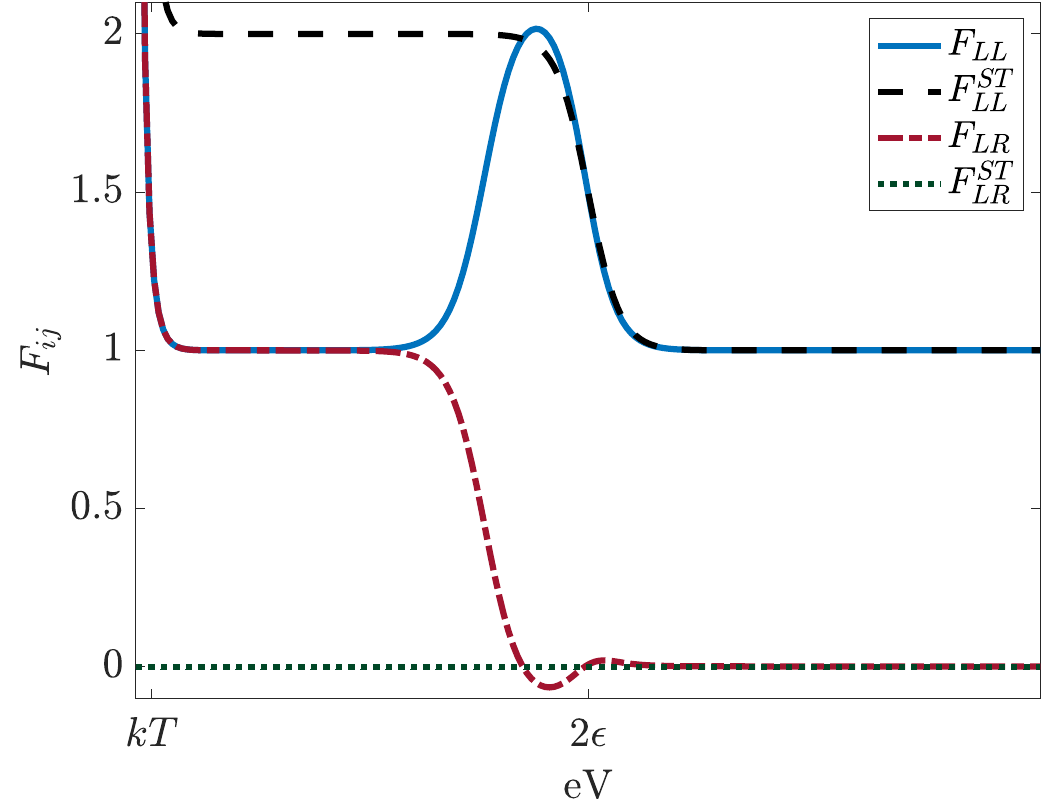}
  \caption{(a) ST and  SOT rates for a symmetrically applied bias voltage, $\mu_L = \mu_R = eV$, shown for small values of $eV < \eps$. Current and noise for the system without QPP, (b) for ST only and (c) for ST and SOT. (d) Corresponding Fano factors. Parameters values used are $k_B T = 0.07\eps$ and $\Gamma_i=7\cdot 10^{-3}\eps$, ($i \in \{L,R\}$).}
    \label{fig_transp_noQP}
\end{figure}

The ST rates from a parity state $|n\rangle$ to a parity state $|m\rangle$ ($m\neq n$) are found to be of the form $W_{mn} = W^{L}_{mn} + W^{R}_{mn}$, with
    \begin{equation}
  W^{R}_{mn} (\chi) =   \Gamma_{R} \left[\left(1-f_{R}^{n,m}\right)e^{-i\chi_R} +f_{R}^{m,n} e^{i\chi_R}\right], \label{eq_Wmn}
\end{equation}
where $\Gamma_R=2\pi\rho_R|t_R|^2$ with $\rho_{R} = (2\pi v_F)^{-1}$ the constant density of states in the right lead and $f_{R}^{n,m} = f_{R}(E_n-E_m) $ is the Fermi function with chemical potential $\mu_{R}$ and temperature $T$ and $W^{L}_{mn} = W^{R}_{mn}(R\leftrightarrow L)$ 
Analytical expressions for the total SOT rates $W_{00}(\chi)$ and $W_{11}(\chi)$ can be found in the Supplemental Material \cite{supplemental}. The rates $W_{nn}(\chi=0)$, and $W_{nm}(\chi=0)$, $n\neq m$, are depicted in Fig.~\ref{fig_transp_noQP}(a). 


The CGF is defined by
$S(\chi,t) = -\ln (\tr (\rho(\chi ,t))) \approx \Lambda_0(\chi)t$ \cite{Bagrets2003,Flindt2010},
where the trace is to be taken over the small system only. 
The last (approximate) equality is valid for long times $t=\tau$, where the eigenvalue with smallest real part, $\Lambda_0$, is adiabatically connected to zero for $\chi \rightarrow 0$.

The mean current and the current noise can be found from the $\chi$-derivatives of the CGF \cite{Flindt2010}, taken at $\chi = 0$ 
\begin{align}
    {I_i} &= i\frac{e}{\tau} \frac{\partial}{\partial \chi_i}S(\chi,\tau) |_{\chi = 0} \\
    S_{ij} & = \frac{e^2}{\tau} \frac{\partial^2}{\partial \chi_i \partial \chi_j}S(\chi,\tau) |_{\chi = 0}.
\end{align}
From these quantities, we calculate the Fano factors $F_{ij}$ as \cite{Fano1947}
\begin{equation}
F_{ij} = \frac{S_{ij}}{e|{I_i}|}.
\end{equation}


Without QPP terms, the parity state $\ket{1}$ is never occupied at low bias voltages, so in this case there is approximately no current from ST processes. When a QPP rate is introduced, the parity is allowed to flip 
without any tunneling process to/from the leads taking place.

To account for such an effective QPP process, we add a phenomenological term to the Liouvillian 
$\mathcal{L} \rightarrow \mathcal{L}_{ST} + \mathcal{L}_{SOT}  +\mathcal{L}_{QP}$.
The Liovilllian $\mathcal{L}_{QP}$ changes the parity in the system and is given by (cf. Fig.~1) \cite{Bittermann2022, Pikulin2012, Cheng2012, SanJose2012, Virtanen2013} 
\begin{equation}
    \mathcal{L}_{QP}[\rho] = -\Gamma_{QP}(c\mathcal{P} \rho \mathcal{P}\cdag + -\frac{1}{2}\{ \rho , \mathcal{P}\cdag c \mathcal{P} \}  ) + h.c.,
\end{equation}
with $\mathcal{P} = (-1)^{c^{\dagger}c}$. For our system we can write this in matrix form as
\begin{equation}
    \mathcal{L}_{QP} = \Gamma_{QP}
    \begin{pmatrix}
     1  & -1 \\
    -1 &  1
    \end{pmatrix}.
    \label{QPP}
\end{equation}

\begin{figure}
  \begin{flushleft}
\large{(a)}
\end{flushleft} 
 \vspace{-0.8cm}
 \centering
 \includegraphics[width=0.42\textwidth, trim=0.8cm 8.2cm 2.45cm 7.5cm, clip]{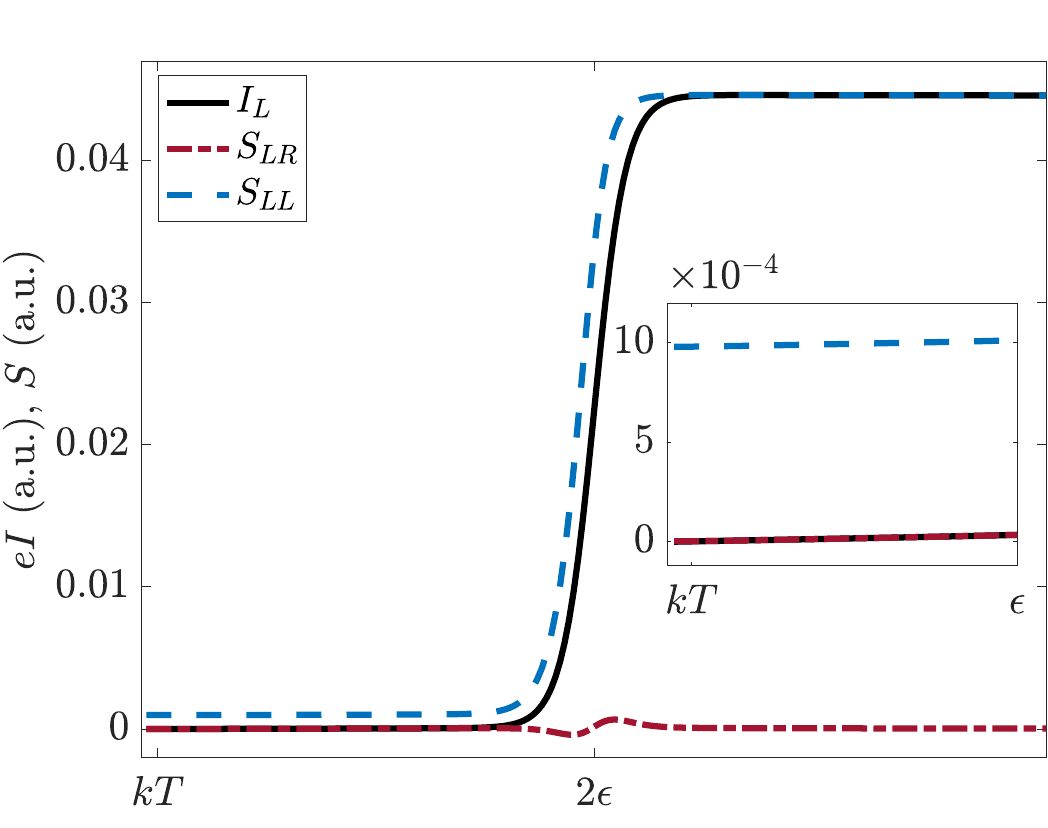}
   \begin{flushleft}
\large{(b)}
\end{flushleft} 
 \vspace{-1.0cm}
       \includegraphics[width=0.42\textwidth, trim=0.9cm 7.5cm 2.35cm 8.2cm, clip]{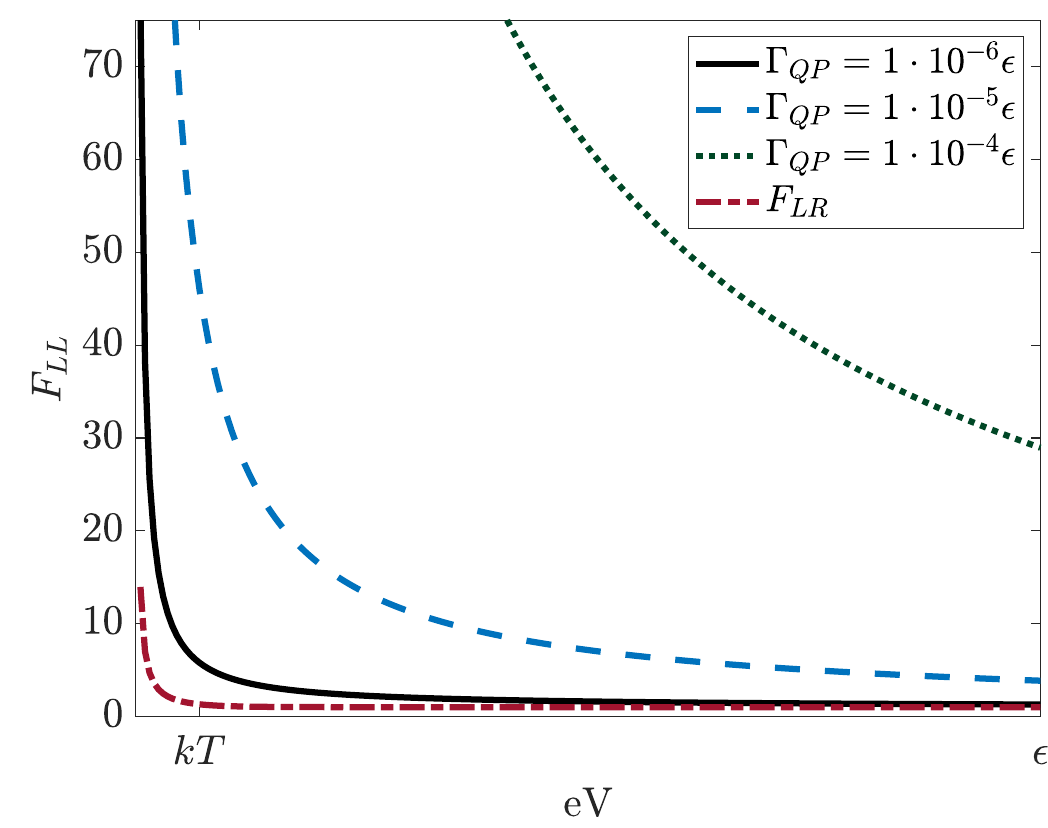}
    \caption{(a) Current and noise as a function of bias voltage for the model including QPP with $\Gamma_{QP}=1\cdot 10^{-4}\eps$ (including both ST and SOT processes). (b) Local noise Fano factor for different values of $\Gamma_{QP}$. The cross noise Fano factor $F_{LR}$ is unaffected by the finite QPP. Parameters used are $k_B T = 0.07\eps$ and $\Gamma_i=7\cdot 10^{-3}\eps$, ($i \in \{L,R\}$).}
    \label{fig_transp_QP}
\end{figure}

\emph{Current and noise without QPP}---
For concreteness, we consider a symmetric bias case $\mu_L=\mu_R=eV>0$ where current flows from the leads to the superconductor \cite{Nilsson2008, Weithofer2014}. The transport properties as a function of $V$ is fundamentally different at low $V$ $(eV\ll 2\epsilon)$ and large $V$ $(eV\gg 2\epsilon)$. At small $V$, the probability to be in $|1\rangle$ is exponentially small at low temperatures and transport is dominated by $H^{SOT}$ only virtually occupying the excited state (the rate is only suppressed by $(1/\epsilon)^2$). 
At large $V$, ST processes are dominant and SOT processes are negligible. It turns out that, as an exact result, ST-events alone cannot lead to cross-correlations, i.e., $S_{LR}^{ST}=0$ (cf. Fig.~\ref{fig_transp_noQP}(b)). However, SOT processes are important at small $V$ (cf. Fig.~\ref{fig_transp_noQP}(c)). As a consequence, the ST-only result contributes wrong Fano factors at small $V$, whereas the result including SOT-contributions agrees quantitatively with Refs.~\onlinecite{Nilsson2008, Weithofer2014} at small and large $V$, see Fig.~\ref{fig_transp_noQP}(d) 
\footnote{The absence of cross correlations $S_{LR}^{ST}=0$ in the ST-only result is correct for any bias profile and a consequence of $\partial_{\chi_i}W_{10}|_{\chi = 0} = \partial_{\chi_i}W_{01}|_{\chi = 0} $ for $i = L,R$, which is always the case when the anomalous and normal tunneling events have equal weights, hence we regard this property as a Majorana-specific result.}.
We note that when $eV\simeq 2\epsilon$, our calculation is not valid since then the SOT-contributions become ill-defined. The uprise of $F_{ij}$ for $eV\lesssim k_BT$ is a consequence of dominant thermal noise contributions.

\emph{Current and noise with QPP}---
In the presence of QPP, the two parity states $|0\rangle$ and $|1\rangle$ become statistically occupied with a switching rate $\Gamma_{QP}$, see Eq.~(\ref{QPP}). We observe that the average current $I_L$ is only weakly affected by QPP, and while the cross-noise $S_{LR}$ is largely unaffected, the local noise $S_{LL}$ is significantly offset to larger values in the presence of QPP (cf. Fig.~\ref{fig_transp_QP}(a)). As a consequence, the local Fano factor $F_{LL}$ becomes much larger than unity and shows a diverging tendency towards small voltages (cf. Fig.~\ref{fig_transp_QP}(b)). Note that 
$S_{LL}$ is significantly enhanced compared to the equilibrium noise when $eV\rightarrow 0$ (compare with $S_{LR}$ and $F_{LR}$ which are largely unaffected by QPP) underlining that QPP is a non-equilibrium effect. These features can be explained considering that QPP induces a significant weight for the system to be in the excited state $|1\rangle$. This opens up the possibility for strong ST-events with rate $W_{01}$ even at small $V$, bringing the system back to the ground state $|0\rangle$. As we can deduce directly from Eq.~(\ref{eq_Wmn}) for $\chi=0$, the ST rate has a normal part $\propto (1-f_{L}(2\epsilon))$ tunneling an electron to the lead, and an anomalous part $\propto f_{L}(-2\epsilon)$ creating a hole in the lead (thereby forming a Cooper pair in the superconducting condensate). Since these processes involve the coupling to a single MBS, the two contributions have the same strength $\Gamma_L$ (and similar for the right lead). This MBS-feature explains the strong $S_{LL}$ (the two parts are added) in the absence of an average current $I_{L}$ (the two parts are subtracted) induced by these strong ST-events. Note that for $eV,k_BT \ll 2\epsilon$, $(1-f_{L,R}(2\epsilon))\approx f_{L,R}(-2\epsilon)\approx 1$ so that the average ST-current is essentially zero and the main contribution to $I_{L}$ originates from crossed Andreev reflection processes (SOT-contributions) \cite{Nilsson2008, Weithofer2014} with $W_{nn}\propto \Gamma_{L}\Gamma_{R}eV/\epsilon^2$ for $k_BT\ll eV$~\cite{supplemental}. In the same limit, we obtain for the Fano factor~\cite{supplemental} $F_{LL}\approx 1+{\bar \rho}_{11}2\pi\epsilon^2/(\Gamma_ReV)$ with ${\bar \rho}_{11}$ the probability to be in state $|1\rangle$ (satisfying ${\cal L}(\chi=0){\bar \rho}=0$). This explains the power-law increase of $F_{LL}$ with decreasing $V$ observed in Fig.~\ref{fig_transp_QP}(b) as well as the fact that QPP only very weakly affects the average current and non-local properties (induced by SOT-processes). We note that in the Majorana-based setups studied in \cite{Colbert2014,Jonckheere2019, Bittermann2022}, $\Gamma_{QP}$ leads to a suppression of shot noise.

In conclusion, we showed that QPP leads to dramatic effects on current noise properties 
of a coupled pair of MBSs weakly tunnel-coupled to current leads including ST- and SOT-events in an open quantum system approach. Particular emphasis is put on the regime of small bias voltage (below the hybridization energy of the two MBSs) applied between the leads and the grounded superconductor. The local Fano factor is shown to grow with decreasing bias voltage in a power-law fashion (reaching values much larger than one), whereas the average current and non-local current fluctuations are largely unaffected by QPP. This striking feature is based on the fact that average current and non-local current fluctuations are dominated by (small) SOT-events, whereas the local noise is dominated by (larger) ST-processes. 
We explain how these QPP-induced noise effects can be traced back to the coupling of each lead to a single MBS. This allows to test characteristic MBS-features via electron transport in the presence of QPP. 


\emph{Acknowledgements}
We thank F.~Dominguez, D.~Loss and S.~Wozny for valuable input and discussions. This research was supported by the Swedish Research Council (VR) and the Deutsche Forschungsgemeinschaft (DFG, German Research Foundation) under Germany’s Excellence Strategy – EXC-2123 QuantumFrontiers – 390837967.

\bibliography{MFs_QPP_paper.bib}

\begin{thebibliography}{67}%
\makeatletter
\providecommand \@ifxundefined [1]{%
 \@ifx{#1\undefined}
}%
\providecommand \@ifnum [1]{%
 \ifnum #1\expandafter \@firstoftwo
 \else \expandafter \@secondoftwo
 \fi
}%
\providecommand \@ifx [1]{%
 \ifx #1\expandafter \@firstoftwo
 \else \expandafter \@secondoftwo
 \fi
}%
\providecommand \natexlab [1]{#1}%
\providecommand \enquote  [1]{``#1''}%
\providecommand \bibnamefont  [1]{#1}%
\providecommand \bibfnamefont [1]{#1}%
\providecommand \citenamefont [1]{#1}%
\providecommand \href@noop [0]{\@secondoftwo}%
\providecommand \href [0]{\begingroup \@sanitize@url \@href}%
\providecommand \@href[1]{\@@startlink{#1}\@@href}%
\providecommand \@@href[1]{\endgroup#1\@@endlink}%
\providecommand \@sanitize@url [0]{\catcode `\\12\catcode `\$12\catcode
  `\&12\catcode `\#12\catcode `\^12\catcode `\_12\catcode `\%12\relax}%
\providecommand \@@startlink[1]{}%
\providecommand \@@endlink[0]{}%
\providecommand \url  [0]{\begingroup\@sanitize@url \@url }%
\providecommand \@url [1]{\endgroup\@href {#1}{\urlprefix }}%
\providecommand \urlprefix  [0]{URL }%
\providecommand \Eprint [0]{\href }%
\providecommand \doibase [0]{http://dx.doi.org/}%
\providecommand \selectlanguage [0]{\@gobble}%
\providecommand \bibinfo  [0]{\@secondoftwo}%
\providecommand \bibfield  [0]{\@secondoftwo}%
\providecommand \translation [1]{[#1]}%
\providecommand \BibitemOpen [0]{}%
\providecommand \bibitemStop [0]{}%
\providecommand \bibitemNoStop [0]{.\EOS\space}%
\providecommand \EOS [0]{\spacefactor3000\relax}%
\providecommand \BibitemShut  [1]{\csname bibitem#1\endcsname}%
\let\auto@bib@innerbib\@empty
\bibitem [{\citenamefont {Read}\ and\ \citenamefont {Green}(2000)}]{Read2000}%
  \BibitemOpen
  \bibfield  {author} {\bibinfo {author} {\bibfnamefont {N.}~\bibnamefont
  {Read}}\ and\ \bibinfo {author} {\bibfnamefont {D.}~\bibnamefont {Green}},\
  }\href {\doibase 10.1103/physrevb.61.10267} {\bibfield  {journal} {\bibinfo
  {journal} {Phys. Rev. B}\ }\textbf {\bibinfo {volume} {61}},\ \bibinfo
  {pages} {10267} (\bibinfo {year} {2000})}\BibitemShut {NoStop}%
\bibitem [{\citenamefont {Kitaev}(2001)}]{Kitaev2001}%
  \BibitemOpen
  \bibfield  {author} {\bibinfo {author} {\bibfnamefont {A.~Y.}\ \bibnamefont
  {Kitaev}},\ }\href {\doibase 10.1070/1063-7869/44/10S/S29} {\bibfield
  {journal} {\bibinfo  {journal} {Physics-Uspekhi}\ }\textbf {\bibinfo {volume}
  {44}},\ \bibinfo {pages} {131} (\bibinfo {year} {2001})}\BibitemShut
  {NoStop}%
\bibitem [{\citenamefont {Fu}\ and\ \citenamefont {Kane}(2008)}]{Fu2008}%
  \BibitemOpen
  \bibfield  {author} {\bibinfo {author} {\bibfnamefont {L.}~\bibnamefont
  {Fu}}\ and\ \bibinfo {author} {\bibfnamefont {C.~L.}\ \bibnamefont {Kane}},\
  }\href {\doibase 10.1103/PhysRevLett.100.096407} {\bibfield  {journal}
  {\bibinfo  {journal} {Phys. Rev. Lett.}\ }\textbf {\bibinfo {volume} {100}},\
  \bibinfo {pages} {096407} (\bibinfo {year} {2008})}\BibitemShut {NoStop}%
\bibitem [{\citenamefont {Oreg}\ \emph {et~al.}(2010)\citenamefont {Oreg},
  \citenamefont {Refael},\ and\ \citenamefont {von Oppen}}]{Oreg2010}%
  \BibitemOpen
  \bibfield  {author} {\bibinfo {author} {\bibfnamefont {Y.}~\bibnamefont
  {Oreg}}, \bibinfo {author} {\bibfnamefont {G.}~\bibnamefont {Refael}}, \ and\
  \bibinfo {author} {\bibfnamefont {F.}~\bibnamefont {von Oppen}},\ }\href
  {\doibase 10.1103/PhysRevLett.105.177002} {\bibfield  {journal} {\bibinfo
  {journal} {Phys. Rev. Lett.}\ }\textbf {\bibinfo {volume} {105}},\ \bibinfo
  {pages} {177002} (\bibinfo {year} {2010})}\BibitemShut {NoStop}%
\bibitem [{\citenamefont {Lutchyn}\ \emph {et~al.}(2010)\citenamefont
  {Lutchyn}, \citenamefont {Sau},\ and\ \citenamefont
  {Das~Sarma}}]{Lutchyn2010}%
  \BibitemOpen
  \bibfield  {author} {\bibinfo {author} {\bibfnamefont {R.~M.}\ \bibnamefont
  {Lutchyn}}, \bibinfo {author} {\bibfnamefont {J.~D.}\ \bibnamefont {Sau}}, \
  and\ \bibinfo {author} {\bibfnamefont {S.}~\bibnamefont {Das~Sarma}},\ }\href
  {\doibase 10.1103/PhysRevLett.105.077001} {\bibfield  {journal} {\bibinfo
  {journal} {Phys. Rev. Lett.}\ }\textbf {\bibinfo {volume} {105}},\ \bibinfo
  {pages} {077001} (\bibinfo {year} {2010})}\BibitemShut {NoStop}%
\bibitem [{\citenamefont {Leijnse}\ and\ \citenamefont
  {Flensberg}(2012)}]{Leijnse2012}%
  \BibitemOpen
  \bibfield  {author} {\bibinfo {author} {\bibfnamefont {M.}~\bibnamefont
  {Leijnse}}\ and\ \bibinfo {author} {\bibfnamefont {K.}~\bibnamefont
  {Flensberg}},\ }\href {\doibase 10.1103/PhysRevB.86.134528} {\bibfield
  {journal} {\bibinfo  {journal} {Phys. Rev. B}\ }\textbf {\bibinfo {volume}
  {86}},\ \bibinfo {pages} {134528} (\bibinfo {year} {2012})}\BibitemShut
  {NoStop}%
\bibitem [{\citenamefont {Beenakker}(2013)}]{Beenakker2013b}%
  \BibitemOpen
  \bibfield  {author} {\bibinfo {author} {\bibfnamefont {C.~W.~J.}\
  \bibnamefont {Beenakker}},\ }\href {\doibase
  10.1146/annurev-conmatphys-030212-184337} {\bibfield  {journal} {\bibinfo
  {journal} {Annual Review of Condensed Matter Physics}\ }\textbf {\bibinfo
  {volume} {4}},\ \bibinfo {pages} {113} (\bibinfo {year} {2013})}\BibitemShut
  {NoStop}%
\bibitem [{\citenamefont {Aguado}(2017)}]{Aguado2017}%
  \BibitemOpen
  \bibfield  {author} {\bibinfo {author} {\bibfnamefont {R.}~\bibnamefont
  {Aguado}},\ }\href {\doibase 10.1393/ncr/i2017-10141-9} {\bibfield  {journal}
  {\bibinfo  {journal} {Nuovo Cimento}\ }\textbf {\bibinfo {volume} {40}},\
  \bibinfo {pages} {523} (\bibinfo {year} {2017})}\BibitemShut {NoStop}%
\bibitem [{\citenamefont {Prada}\ \emph {et~al.}(2020)\citenamefont {Prada},
  \citenamefont {San-Jose}, \citenamefont {de~Moor}, \citenamefont {Geresdi},
  \citenamefont {Lee}, \citenamefont {Klinovaja}, \citenamefont {Loss},
  \citenamefont {Nygard}, \citenamefont {Aguado},\ and\ \citenamefont
  {Kouwenhoven}}]{Prada2020a}%
  \BibitemOpen
  \bibfield  {author} {\bibinfo {author} {\bibfnamefont {E.}~\bibnamefont
  {Prada}}, \bibinfo {author} {\bibfnamefont {P.}~\bibnamefont {San-Jose}},
  \bibinfo {author} {\bibfnamefont {M.~W.~A.}\ \bibnamefont {de~Moor}},
  \bibinfo {author} {\bibfnamefont {A.}~\bibnamefont {Geresdi}}, \bibinfo
  {author} {\bibfnamefont {E.~J.~H.}\ \bibnamefont {Lee}}, \bibinfo {author}
  {\bibfnamefont {J.}~\bibnamefont {Klinovaja}}, \bibinfo {author}
  {\bibfnamefont {D.}~\bibnamefont {Loss}}, \bibinfo {author} {\bibfnamefont
  {J.}~\bibnamefont {Nygard}}, \bibinfo {author} {\bibfnamefont
  {R.}~\bibnamefont {Aguado}}, \ and\ \bibinfo {author} {\bibfnamefont {L.~P.}\
  \bibnamefont {Kouwenhoven}},\ }\href {\doibase 10.1038/s42254-020-0228-y}
  {\bibfield  {journal} {\bibinfo  {journal} {Nat. Rev. Phys.}\ }\textbf
  {\bibinfo {volume} {2}},\ \bibinfo {pages} {575} (\bibinfo {year}
  {2020})}\BibitemShut {NoStop}%
\bibitem [{\citenamefont {Nayak}\ \emph {et~al.}(2008)\citenamefont {Nayak},
  \citenamefont {Simon}, \citenamefont {Stern}, \citenamefont {Freedman},\ and\
  \citenamefont {Das~Sarma}}]{Nayak2008}%
  \BibitemOpen
  \bibfield  {author} {\bibinfo {author} {\bibfnamefont {C.}~\bibnamefont
  {Nayak}}, \bibinfo {author} {\bibfnamefont {S.~H.}\ \bibnamefont {Simon}},
  \bibinfo {author} {\bibfnamefont {A.}~\bibnamefont {Stern}}, \bibinfo
  {author} {\bibfnamefont {M.}~\bibnamefont {Freedman}}, \ and\ \bibinfo
  {author} {\bibfnamefont {S.}~\bibnamefont {Das~Sarma}},\ }\href {\doibase
  10.1103/RevModPhys.80.1083} {\bibfield  {journal} {\bibinfo  {journal} {Rev.
  Mod. Phys.}\ }\textbf {\bibinfo {volume} {80}},\ \bibinfo {pages} {1083}
  (\bibinfo {year} {2008})}\BibitemShut {NoStop}%
\bibitem [{\citenamefont {Kitaev}(2003)}]{Kitaev2003}%
  \BibitemOpen
  \bibfield  {author} {\bibinfo {author} {\bibfnamefont {A.}~\bibnamefont
  {Kitaev}},\ }\href {\doibase https://doi.org/10.1016/S0003-4916(02)00018-0}
  {\bibfield  {journal} {\bibinfo  {journal} {Annals of Physics}\ }\textbf
  {\bibinfo {volume} {303}},\ \bibinfo {pages} {2} (\bibinfo {year}
  {2003})}\BibitemShut {NoStop}%
\bibitem [{\citenamefont {Aasen}\ \emph {et~al.}(2016)\citenamefont {Aasen},
  \citenamefont {Hell}, \citenamefont {Mishmash}, \citenamefont {Higginbotham},
  \citenamefont {Danon}, \citenamefont {Leijnse}, \citenamefont {Jespersen},
  \citenamefont {Folk}, \citenamefont {Marcus}, \citenamefont {Flensberg},\
  and\ \citenamefont {Alicea}}]{Aasen2016}%
  \BibitemOpen
  \bibfield  {author} {\bibinfo {author} {\bibfnamefont {D.}~\bibnamefont
  {Aasen}}, \bibinfo {author} {\bibfnamefont {M.}~\bibnamefont {Hell}},
  \bibinfo {author} {\bibfnamefont {R.~V.}\ \bibnamefont {Mishmash}}, \bibinfo
  {author} {\bibfnamefont {A.}~\bibnamefont {Higginbotham}}, \bibinfo {author}
  {\bibfnamefont {J.}~\bibnamefont {Danon}}, \bibinfo {author} {\bibfnamefont
  {M.}~\bibnamefont {Leijnse}}, \bibinfo {author} {\bibfnamefont {T.~S.}\
  \bibnamefont {Jespersen}}, \bibinfo {author} {\bibfnamefont {J.~A.}\
  \bibnamefont {Folk}}, \bibinfo {author} {\bibfnamefont {C.~M.}\ \bibnamefont
  {Marcus}}, \bibinfo {author} {\bibfnamefont {K.}~\bibnamefont {Flensberg}}, \
  and\ \bibinfo {author} {\bibfnamefont {J.}~\bibnamefont {Alicea}},\ }\href
  {\doibase 10.1103/PhysRevX.6.031016} {\bibfield  {journal} {\bibinfo
  {journal} {Phys. Rev. X}\ }\textbf {\bibinfo {volume} {6}},\ \bibinfo {pages}
  {031016} (\bibinfo {year} {2016})}\BibitemShut {NoStop}%
\bibitem [{\citenamefont {Hoffman}\ \emph {et~al.}(2016)\citenamefont
  {Hoffman}, \citenamefont {Schrade}, \citenamefont {Klinovaja},\ and\
  \citenamefont {Loss}}]{Hoffman2016}%
  \BibitemOpen
  \bibfield  {author} {\bibinfo {author} {\bibfnamefont {S.}~\bibnamefont
  {Hoffman}}, \bibinfo {author} {\bibfnamefont {C.}~\bibnamefont {Schrade}},
  \bibinfo {author} {\bibfnamefont {J.}~\bibnamefont {Klinovaja}}, \ and\
  \bibinfo {author} {\bibfnamefont {D.}~\bibnamefont {Loss}},\ }\href {\doibase
  10.1103/PhysRevB.94.045316} {\bibfield  {journal} {\bibinfo  {journal} {Phys.
  Rev. B}\ }\textbf {\bibinfo {volume} {94}},\ \bibinfo {pages} {045316}
  (\bibinfo {year} {2016})}\BibitemShut {NoStop}%
\bibitem [{\citenamefont {Landau}\ \emph {et~al.}(2016)\citenamefont {Landau},
  \citenamefont {Plugge}, \citenamefont {Sela}, \citenamefont {Altland},
  \citenamefont {Albrecht},\ and\ \citenamefont {Egger}}]{Landau2016}%
  \BibitemOpen
  \bibfield  {author} {\bibinfo {author} {\bibfnamefont {L.~A.}\ \bibnamefont
  {Landau}}, \bibinfo {author} {\bibfnamefont {S.}~\bibnamefont {Plugge}},
  \bibinfo {author} {\bibfnamefont {E.}~\bibnamefont {Sela}}, \bibinfo {author}
  {\bibfnamefont {A.}~\bibnamefont {Altland}}, \bibinfo {author} {\bibfnamefont
  {S.~M.}\ \bibnamefont {Albrecht}}, \ and\ \bibinfo {author} {\bibfnamefont
  {R.}~\bibnamefont {Egger}},\ }\href {\doibase 10.1103/PhysRevLett.116.050501}
  {\bibfield  {journal} {\bibinfo  {journal} {Phys. Rev. Lett.}\ }\textbf
  {\bibinfo {volume} {116}},\ \bibinfo {pages} {050501} (\bibinfo {year}
  {2016})}\BibitemShut {NoStop}%
\bibitem [{\citenamefont {Plugge}\ \emph {et~al.}(2017)\citenamefont {Plugge},
  \citenamefont {Rasmussen}, \citenamefont {Egger},\ and\ \citenamefont
  {Flensberg}}]{Plugge2017}%
  \BibitemOpen
  \bibfield  {author} {\bibinfo {author} {\bibfnamefont {S.}~\bibnamefont
  {Plugge}}, \bibinfo {author} {\bibfnamefont {A.}~\bibnamefont {Rasmussen}},
  \bibinfo {author} {\bibfnamefont {R.}~\bibnamefont {Egger}}, \ and\ \bibinfo
  {author} {\bibfnamefont {K.}~\bibnamefont {Flensberg}},\ }\href {\doibase
  10.1088/1367-2630/aa54e1} {\bibfield  {journal} {\bibinfo  {journal} {New
  Journal of Physics}\ }\textbf {\bibinfo {volume} {19}},\ \bibinfo {pages}
  {012001} (\bibinfo {year} {2017})}\BibitemShut {NoStop}%
\bibitem [{\citenamefont {Park}\ \emph {et~al.}(2020)\citenamefont {Park},
  \citenamefont {Sim},\ and\ \citenamefont {Recher}}]{Park2020}%
  \BibitemOpen
  \bibfield  {author} {\bibinfo {author} {\bibfnamefont {S.}~\bibnamefont
  {Park}}, \bibinfo {author} {\bibfnamefont {H.-S.}\ \bibnamefont {Sim}}, \
  and\ \bibinfo {author} {\bibfnamefont {P.}~\bibnamefont {Recher}},\ }\href
  {\doibase 10.1103/PhysRevLett.125.187702} {\bibfield  {journal} {\bibinfo
  {journal} {Phys. Rev. Lett.}\ }\textbf {\bibinfo {volume} {125}},\ \bibinfo
  {pages} {187702} (\bibinfo {year} {2020})}\BibitemShut {NoStop}%
\bibitem [{\citenamefont {Goldstein}\ and\ \citenamefont
  {Chamon}(2011)}]{Goldstein2011}%
  \BibitemOpen
  \bibfield  {author} {\bibinfo {author} {\bibfnamefont {G.}~\bibnamefont
  {Goldstein}}\ and\ \bibinfo {author} {\bibfnamefont {C.}~\bibnamefont
  {Chamon}},\ }\href {\doibase 10.1103/PhysRevB.84.205109} {\bibfield
  {journal} {\bibinfo  {journal} {Phys. Rev. B}\ }\textbf {\bibinfo {volume}
  {84}},\ \bibinfo {pages} {205109} (\bibinfo {year} {2011})}\BibitemShut
  {NoStop}%
\bibitem [{\citenamefont {Budich}\ \emph {et~al.}(2012)\citenamefont {Budich},
  \citenamefont {Walter},\ and\ \citenamefont {Trauzettel}}]{Budich2012}%
  \BibitemOpen
  \bibfield  {author} {\bibinfo {author} {\bibfnamefont {J.~C.}\ \bibnamefont
  {Budich}}, \bibinfo {author} {\bibfnamefont {S.}~\bibnamefont {Walter}}, \
  and\ \bibinfo {author} {\bibfnamefont {B.}~\bibnamefont {Trauzettel}},\
  }\href {\doibase 10.1103/PhysRevB.85.121405} {\bibfield  {journal} {\bibinfo
  {journal} {Phys. Rev. B}\ }\textbf {\bibinfo {volume} {85}},\ \bibinfo
  {pages} {121405} (\bibinfo {year} {2012})}\BibitemShut {NoStop}%
\bibitem [{\citenamefont {Rainis}\ and\ \citenamefont
  {Loss}(2012)}]{Rainis2012}%
  \BibitemOpen
  \bibfield  {author} {\bibinfo {author} {\bibfnamefont {D.}~\bibnamefont
  {Rainis}}\ and\ \bibinfo {author} {\bibfnamefont {D.}~\bibnamefont {Loss}},\
  }\href {\doibase 10.1103/PhysRevB.85.174533} {\bibfield  {journal} {\bibinfo
  {journal} {Phys. Rev. B}\ }\textbf {\bibinfo {volume} {85}},\ \bibinfo
  {pages} {174533} (\bibinfo {year} {2012})}\BibitemShut {NoStop}%
\bibitem [{\citenamefont {Plugge}\ \emph {et~al.}(2016)\citenamefont {Plugge},
  \citenamefont {Zazunov}, \citenamefont {Eriksson}, \citenamefont {Tsvelik},\
  and\ \citenamefont {Egger}}]{Plugge2016}%
  \BibitemOpen
  \bibfield  {author} {\bibinfo {author} {\bibfnamefont {S.}~\bibnamefont
  {Plugge}}, \bibinfo {author} {\bibfnamefont {A.}~\bibnamefont {Zazunov}},
  \bibinfo {author} {\bibfnamefont {E.}~\bibnamefont {Eriksson}}, \bibinfo
  {author} {\bibfnamefont {A.~M.}\ \bibnamefont {Tsvelik}}, \ and\ \bibinfo
  {author} {\bibfnamefont {R.}~\bibnamefont {Egger}},\ }\href {\doibase
  10.1103/PhysRevB.93.104524} {\bibfield  {journal} {\bibinfo  {journal} {Phys.
  Rev. B}\ }\textbf {\bibinfo {volume} {93}},\ \bibinfo {pages} {104524}
  (\bibinfo {year} {2016})}\BibitemShut {NoStop}%
\bibitem [{\citenamefont {Albrecht}\ \emph {et~al.}(2017)\citenamefont
  {Albrecht}, \citenamefont {Hansen}, \citenamefont {Higginbotham},
  \citenamefont {Kuemmeth}, \citenamefont {Jespersen}, \citenamefont
  {Nyg\aa{}rd}, \citenamefont {Krogstrup}, \citenamefont {Danon}, \citenamefont
  {Flensberg},\ and\ \citenamefont {Marcus}}]{Albrecht2017}%
  \BibitemOpen
  \bibfield  {author} {\bibinfo {author} {\bibfnamefont {S.~M.}\ \bibnamefont
  {Albrecht}}, \bibinfo {author} {\bibfnamefont {E.~B.}\ \bibnamefont
  {Hansen}}, \bibinfo {author} {\bibfnamefont {A.~P.}\ \bibnamefont
  {Higginbotham}}, \bibinfo {author} {\bibfnamefont {F.}~\bibnamefont
  {Kuemmeth}}, \bibinfo {author} {\bibfnamefont {T.~S.}\ \bibnamefont
  {Jespersen}}, \bibinfo {author} {\bibfnamefont {J.}~\bibnamefont
  {Nyg\aa{}rd}}, \bibinfo {author} {\bibfnamefont {P.}~\bibnamefont
  {Krogstrup}}, \bibinfo {author} {\bibfnamefont {J.}~\bibnamefont {Danon}},
  \bibinfo {author} {\bibfnamefont {K.}~\bibnamefont {Flensberg}}, \ and\
  \bibinfo {author} {\bibfnamefont {C.~M.}\ \bibnamefont {Marcus}},\ }\href
  {\doibase 10.1103/PhysRevLett.118.137701} {\bibfield  {journal} {\bibinfo
  {journal} {Phys. Rev. Lett.}\ }\textbf {\bibinfo {volume} {118}},\ \bibinfo
  {pages} {137701} (\bibinfo {year} {2017})}\BibitemShut {NoStop}%
\bibitem [{\citenamefont {Schulenborg}\ \emph {et~al.}(2023)\citenamefont
  {Schulenborg}, \citenamefont {Kr\o{}jer}, \citenamefont {Burrello},
  \citenamefont {Leijnse},\ and\ \citenamefont {Flensberg}}]{Schulenborg2023}%
  \BibitemOpen
  \bibfield  {author} {\bibinfo {author} {\bibfnamefont {J.}~\bibnamefont
  {Schulenborg}}, \bibinfo {author} {\bibfnamefont {S.}~\bibnamefont
  {Kr\o{}jer}}, \bibinfo {author} {\bibfnamefont {M.}~\bibnamefont {Burrello}},
  \bibinfo {author} {\bibfnamefont {M.}~\bibnamefont {Leijnse}}, \ and\
  \bibinfo {author} {\bibfnamefont {K.}~\bibnamefont {Flensberg}},\ }\href
  {\doibase 10.1103/PhysRevB.107.L121401} {\bibfield  {journal} {\bibinfo
  {journal} {Phys. Rev. B}\ }\textbf {\bibinfo {volume} {107}},\ \bibinfo
  {pages} {L121401} (\bibinfo {year} {2023})}\BibitemShut {NoStop}%
\bibitem [{\citenamefont {Aumentado}\ \emph {et~al.}(2004)\citenamefont
  {Aumentado}, \citenamefont {Keller}, \citenamefont {Martinis},\ and\
  \citenamefont {Devoret}}]{Aumentado2004}%
  \BibitemOpen
  \bibfield  {author} {\bibinfo {author} {\bibfnamefont {J.}~\bibnamefont
  {Aumentado}}, \bibinfo {author} {\bibfnamefont {M.~W.}\ \bibnamefont
  {Keller}}, \bibinfo {author} {\bibfnamefont {J.~M.}\ \bibnamefont
  {Martinis}}, \ and\ \bibinfo {author} {\bibfnamefont {M.~H.}\ \bibnamefont
  {Devoret}},\ }\href {\doibase 10.1103/PhysRevLett.92.066802} {\bibfield
  {journal} {\bibinfo  {journal} {Phys. Rev. Lett.}\ }\textbf {\bibinfo
  {volume} {92}},\ \bibinfo {pages} {066802} (\bibinfo {year}
  {2004})}\BibitemShut {NoStop}%
\bibitem [{\citenamefont {Lutchyn}\ \emph {et~al.}(2005)\citenamefont
  {Lutchyn}, \citenamefont {Glazman},\ and\ \citenamefont
  {Larkin}}]{Lutchyn2005}%
  \BibitemOpen
  \bibfield  {author} {\bibinfo {author} {\bibfnamefont {R.}~\bibnamefont
  {Lutchyn}}, \bibinfo {author} {\bibfnamefont {L.}~\bibnamefont {Glazman}}, \
  and\ \bibinfo {author} {\bibfnamefont {A.}~\bibnamefont {Larkin}},\ }\href
  {\doibase 10.1103/PhysRevB.72.014517} {\bibfield  {journal} {\bibinfo
  {journal} {Phys. Rev. B}\ }\textbf {\bibinfo {volume} {72}},\ \bibinfo
  {pages} {014517} (\bibinfo {year} {2005})}\BibitemShut {NoStop}%
\bibitem [{\citenamefont {Catelani}\ \emph {et~al.}(2011)\citenamefont
  {Catelani}, \citenamefont {Koch}, \citenamefont {Frunzio}, \citenamefont
  {Schoelkopf}, \citenamefont {Devoret},\ and\ \citenamefont
  {Glazman}}]{Catelani2011}%
  \BibitemOpen
  \bibfield  {author} {\bibinfo {author} {\bibfnamefont {G.}~\bibnamefont
  {Catelani}}, \bibinfo {author} {\bibfnamefont {J.}~\bibnamefont {Koch}},
  \bibinfo {author} {\bibfnamefont {L.}~\bibnamefont {Frunzio}}, \bibinfo
  {author} {\bibfnamefont {R.~J.}\ \bibnamefont {Schoelkopf}}, \bibinfo
  {author} {\bibfnamefont {M.~H.}\ \bibnamefont {Devoret}}, \ and\ \bibinfo
  {author} {\bibfnamefont {L.~I.}\ \bibnamefont {Glazman}},\ }\href {\doibase
  10.1103/PhysRevLett.106.077002} {\bibfield  {journal} {\bibinfo  {journal}
  {Phys. Rev. Lett.}\ }\textbf {\bibinfo {volume} {106}},\ \bibinfo {pages}
  {077002} (\bibinfo {year} {2011})}\BibitemShut {NoStop}%
\bibitem [{\citenamefont {Liu}\ \emph {et~al.}(2024)\citenamefont {Liu},
  \citenamefont {Harrison}, \citenamefont {Patel}, \citenamefont {Wilen},
  \citenamefont {Rafferty}, \citenamefont {Shearrow}, \citenamefont {Ballard},
  \citenamefont {Iaia}, \citenamefont {Ku}, \citenamefont {Plourde},\ and\
  \citenamefont {McDermott}}]{Liu2024}%
  \BibitemOpen
  \bibfield  {author} {\bibinfo {author} {\bibfnamefont {C.~H.}\ \bibnamefont
  {Liu}}, \bibinfo {author} {\bibfnamefont {D.~C.}\ \bibnamefont {Harrison}},
  \bibinfo {author} {\bibfnamefont {S.}~\bibnamefont {Patel}}, \bibinfo
  {author} {\bibfnamefont {C.~D.}\ \bibnamefont {Wilen}}, \bibinfo {author}
  {\bibfnamefont {O.}~\bibnamefont {Rafferty}}, \bibinfo {author}
  {\bibfnamefont {A.}~\bibnamefont {Shearrow}}, \bibinfo {author}
  {\bibfnamefont {A.}~\bibnamefont {Ballard}}, \bibinfo {author} {\bibfnamefont
  {V.}~\bibnamefont {Iaia}}, \bibinfo {author} {\bibfnamefont {J.}~\bibnamefont
  {Ku}}, \bibinfo {author} {\bibfnamefont {B.~L.~T.}\ \bibnamefont {Plourde}},
  \ and\ \bibinfo {author} {\bibfnamefont {R.}~\bibnamefont {McDermott}},\
  }\href {\doibase 10.1103/PhysRevLett.132.017001} {\bibfield  {journal}
  {\bibinfo  {journal} {Phys. Rev. Lett.}\ }\textbf {\bibinfo {volume} {132}},\
  \bibinfo {pages} {017001} (\bibinfo {year} {2024})}\BibitemShut {NoStop}%
\bibitem [{\citenamefont {Tsintzis}\ \emph {et~al.}(2024)\citenamefont
  {Tsintzis}, \citenamefont {Souto}, \citenamefont {Flensberg}, \citenamefont
  {Danon},\ and\ \citenamefont {Leijnse}}]{Tsintzis2024}%
  \BibitemOpen
  \bibfield  {author} {\bibinfo {author} {\bibfnamefont {A.}~\bibnamefont
  {Tsintzis}}, \bibinfo {author} {\bibfnamefont {R.~S.}\ \bibnamefont {Souto}},
  \bibinfo {author} {\bibfnamefont {K.}~\bibnamefont {Flensberg}}, \bibinfo
  {author} {\bibfnamefont {J.}~\bibnamefont {Danon}}, \ and\ \bibinfo {author}
  {\bibfnamefont {M.}~\bibnamefont {Leijnse}},\ }\href {\doibase
  10.1103/PRXQuantum.5.010323} {\bibfield  {journal} {\bibinfo  {journal} {PRX
  Quantum}\ }\textbf {\bibinfo {volume} {5}},\ \bibinfo {pages} {010323}
  (\bibinfo {year} {2024})}\BibitemShut {NoStop}%
\bibitem [{\citenamefont {Bolech}\ and\ \citenamefont
  {Demler}(2007)}]{Bolech2007}%
  \BibitemOpen
  \bibfield  {author} {\bibinfo {author} {\bibfnamefont {C.~J.}\ \bibnamefont
  {Bolech}}\ and\ \bibinfo {author} {\bibfnamefont {E.}~\bibnamefont
  {Demler}},\ }\href {\doibase 10.1103/PhysRevLett.98.237002} {\bibfield
  {journal} {\bibinfo  {journal} {Phys. Rev. Lett.}\ }\textbf {\bibinfo
  {volume} {98}},\ \bibinfo {pages} {237002} (\bibinfo {year}
  {2007})}\BibitemShut {NoStop}%
\bibitem [{\citenamefont {Nilsson}\ \emph {et~al.}(2008)\citenamefont
  {Nilsson}, \citenamefont {Akhmerov},\ and\ \citenamefont
  {Beenakker}}]{Nilsson2008}%
  \BibitemOpen
  \bibfield  {author} {\bibinfo {author} {\bibfnamefont {J.}~\bibnamefont
  {Nilsson}}, \bibinfo {author} {\bibfnamefont {A.~R.}\ \bibnamefont
  {Akhmerov}}, \ and\ \bibinfo {author} {\bibfnamefont {C.~W.~J.}\ \bibnamefont
  {Beenakker}},\ }\href {\doibase 10.1103/PhysRevLett.101.120403} {\bibfield
  {journal} {\bibinfo  {journal} {Phys. Rev. Lett.}\ }\textbf {\bibinfo
  {volume} {101}},\ \bibinfo {pages} {120403} (\bibinfo {year}
  {2008})}\BibitemShut {NoStop}%
\bibitem [{\citenamefont {Prada}\ \emph {et~al.}(2012)\citenamefont {Prada},
  \citenamefont {San-Jose},\ and\ \citenamefont {Aguado}}]{Prada2012}%
  \BibitemOpen
  \bibfield  {author} {\bibinfo {author} {\bibfnamefont {E.}~\bibnamefont
  {Prada}}, \bibinfo {author} {\bibfnamefont {P.}~\bibnamefont {San-Jose}}, \
  and\ \bibinfo {author} {\bibfnamefont {R.}~\bibnamefont {Aguado}},\ }\href
  {\doibase 10.1103/PhysRevB.86.180503} {\bibfield  {journal} {\bibinfo
  {journal} {Phys. Rev. B}\ }\textbf {\bibinfo {volume} {86}},\ \bibinfo
  {pages} {180503} (\bibinfo {year} {2012})}\BibitemShut {NoStop}%
\bibitem [{\citenamefont {Wu}\ and\ \citenamefont {Cao}(2012)}]{Wu2012}%
  \BibitemOpen
  \bibfield  {author} {\bibinfo {author} {\bibfnamefont {B.~H.}\ \bibnamefont
  {Wu}}\ and\ \bibinfo {author} {\bibfnamefont {J.~C.}\ \bibnamefont {Cao}},\
  }\href {\doibase 10.1103/PhysRevB.85.085415} {\bibfield  {journal} {\bibinfo
  {journal} {Phys. Rev. B}\ }\textbf {\bibinfo {volume} {85}},\ \bibinfo
  {pages} {085415} (\bibinfo {year} {2012})}\BibitemShut {NoStop}%
\bibitem [{\citenamefont {H\"utzen}\ \emph {et~al.}(2012)\citenamefont
  {H\"utzen}, \citenamefont {Zazunov}, \citenamefont {Braunecker},
  \citenamefont {Yeyati},\ and\ \citenamefont {Egger}}]{Huetzen2012}%
  \BibitemOpen
  \bibfield  {author} {\bibinfo {author} {\bibfnamefont {R.}~\bibnamefont
  {H\"utzen}}, \bibinfo {author} {\bibfnamefont {A.}~\bibnamefont {Zazunov}},
  \bibinfo {author} {\bibfnamefont {B.}~\bibnamefont {Braunecker}}, \bibinfo
  {author} {\bibfnamefont {A.~L.}\ \bibnamefont {Yeyati}}, \ and\ \bibinfo
  {author} {\bibfnamefont {R.}~\bibnamefont {Egger}},\ }\href {\doibase
  10.1103/PhysRevLett.109.166403} {\bibfield  {journal} {\bibinfo  {journal}
  {Phys. Rev. Lett.}\ }\textbf {\bibinfo {volume} {109}},\ \bibinfo {pages}
  {166403} (\bibinfo {year} {2012})}\BibitemShut {NoStop}%
\bibitem [{\citenamefont {Rainis}\ \emph {et~al.}(2013)\citenamefont {Rainis},
  \citenamefont {Trifunovic}, \citenamefont {Klinovaja},\ and\ \citenamefont
  {Loss}}]{Rainis2013}%
  \BibitemOpen
  \bibfield  {author} {\bibinfo {author} {\bibfnamefont {D.}~\bibnamefont
  {Rainis}}, \bibinfo {author} {\bibfnamefont {L.}~\bibnamefont {Trifunovic}},
  \bibinfo {author} {\bibfnamefont {J.}~\bibnamefont {Klinovaja}}, \ and\
  \bibinfo {author} {\bibfnamefont {D.}~\bibnamefont {Loss}},\ }\href {\doibase
  10.1103/PhysRevB.87.024515} {\bibfield  {journal} {\bibinfo  {journal} {Phys.
  Rev. B}\ }\textbf {\bibinfo {volume} {87}},\ \bibinfo {pages} {024515}
  (\bibinfo {year} {2013})}\BibitemShut {NoStop}%
\bibitem [{\citenamefont {Zocher}\ and\ \citenamefont
  {Rosenow}(2013)}]{Zocher2013}%
  \BibitemOpen
  \bibfield  {author} {\bibinfo {author} {\bibfnamefont {B.}~\bibnamefont
  {Zocher}}\ and\ \bibinfo {author} {\bibfnamefont {B.}~\bibnamefont
  {Rosenow}},\ }\href {\doibase 10.1103/PhysRevLett.111.036802} {\bibfield
  {journal} {\bibinfo  {journal} {Phys. Rev. Lett.}\ }\textbf {\bibinfo
  {volume} {111}},\ \bibinfo {pages} {036802} (\bibinfo {year}
  {2013})}\BibitemShut {NoStop}%
\bibitem [{\citenamefont {Weithofer}\ \emph {et~al.}(2014)\citenamefont
  {Weithofer}, \citenamefont {Recher},\ and\ \citenamefont
  {Schmidt}}]{Weithofer2014}%
  \BibitemOpen
  \bibfield  {author} {\bibinfo {author} {\bibfnamefont {L.}~\bibnamefont
  {Weithofer}}, \bibinfo {author} {\bibfnamefont {P.}~\bibnamefont {Recher}}, \
  and\ \bibinfo {author} {\bibfnamefont {T.~L.}\ \bibnamefont {Schmidt}},\
  }\href {\doibase 10.1103/PhysRevB.90.205416} {\bibfield  {journal} {\bibinfo
  {journal} {Phys. Rev. B}\ }\textbf {\bibinfo {volume} {90}},\ \bibinfo
  {pages} {205416} (\bibinfo {year} {2014})}\BibitemShut {NoStop}%
\bibitem [{\citenamefont {van Heck}\ \emph {et~al.}(2016)\citenamefont {van
  Heck}, \citenamefont {Lutchyn},\ and\ \citenamefont {Glazman}}]{vanHeck2016}%
  \BibitemOpen
  \bibfield  {author} {\bibinfo {author} {\bibfnamefont {B.}~\bibnamefont {van
  Heck}}, \bibinfo {author} {\bibfnamefont {R.~M.}\ \bibnamefont {Lutchyn}}, \
  and\ \bibinfo {author} {\bibfnamefont {L.~I.}\ \bibnamefont {Glazman}},\
  }\href {\doibase 10.1103/PhysRevB.93.235431} {\bibfield  {journal} {\bibinfo
  {journal} {Phys. Rev. B}\ }\textbf {\bibinfo {volume} {93}},\ \bibinfo
  {pages} {235431} (\bibinfo {year} {2016})}\BibitemShut {NoStop}%
\bibitem [{\citenamefont {Zazunov}\ \emph {et~al.}(2016)\citenamefont
  {Zazunov}, \citenamefont {Egger},\ and\ \citenamefont
  {Levy~Yeyati}}]{Zazunov2016}%
  \BibitemOpen
  \bibfield  {author} {\bibinfo {author} {\bibfnamefont {A.}~\bibnamefont
  {Zazunov}}, \bibinfo {author} {\bibfnamefont {R.}~\bibnamefont {Egger}}, \
  and\ \bibinfo {author} {\bibfnamefont {A.}~\bibnamefont {Levy~Yeyati}},\
  }\href {\doibase 10.1103/PhysRevB.94.014502} {\bibfield  {journal} {\bibinfo
  {journal} {Phys. Rev. B}\ }\textbf {\bibinfo {volume} {94}},\ \bibinfo
  {pages} {014502} (\bibinfo {year} {2016})}\BibitemShut {NoStop}%
\bibitem [{\citenamefont {Liu}\ \emph {et~al.}(2017)\citenamefont {Liu},
  \citenamefont {Sau}, \citenamefont {Stanescu},\ and\ \citenamefont
  {Das~Sarma}}]{Liu2017}%
  \BibitemOpen
  \bibfield  {author} {\bibinfo {author} {\bibfnamefont {C.-X.}\ \bibnamefont
  {Liu}}, \bibinfo {author} {\bibfnamefont {J.~D.}\ \bibnamefont {Sau}},
  \bibinfo {author} {\bibfnamefont {T.~D.}\ \bibnamefont {Stanescu}}, \ and\
  \bibinfo {author} {\bibfnamefont {S.}~\bibnamefont {Das~Sarma}},\ }\href
  {\doibase 10.1103/PhysRevB.96.075161} {\bibfield  {journal} {\bibinfo
  {journal} {Phys. Rev. B}\ }\textbf {\bibinfo {volume} {96}},\ \bibinfo
  {pages} {075161} (\bibinfo {year} {2017})}\BibitemShut {NoStop}%
\bibitem [{\citenamefont {Schuray}\ \emph {et~al.}(2017)\citenamefont
  {Schuray}, \citenamefont {Weithofer},\ and\ \citenamefont
  {Recher}}]{Schuray2017}%
  \BibitemOpen
  \bibfield  {author} {\bibinfo {author} {\bibfnamefont {A.}~\bibnamefont
  {Schuray}}, \bibinfo {author} {\bibfnamefont {L.}~\bibnamefont {Weithofer}},
  \ and\ \bibinfo {author} {\bibfnamefont {P.}~\bibnamefont {Recher}},\ }\href
  {\doibase 10.1103/PhysRevB.96.085417} {\bibfield  {journal} {\bibinfo
  {journal} {Phys. Rev. B}\ }\textbf {\bibinfo {volume} {96}},\ \bibinfo
  {pages} {085417} (\bibinfo {year} {2017})}\BibitemShut {NoStop}%
\bibitem [{\citenamefont {Prada}\ \emph {et~al.}(2017)\citenamefont {Prada},
  \citenamefont {Aguado},\ and\ \citenamefont {San-Jose}}]{Prada2017a}%
  \BibitemOpen
  \bibfield  {author} {\bibinfo {author} {\bibfnamefont {E.}~\bibnamefont
  {Prada}}, \bibinfo {author} {\bibfnamefont {R.}~\bibnamefont {Aguado}}, \
  and\ \bibinfo {author} {\bibfnamefont {P.}~\bibnamefont {San-Jose}},\ }\href
  {\doibase 10.1103/PhysRevB.96.085418} {\bibfield  {journal} {\bibinfo
  {journal} {Phys. Rev. B}\ }\textbf {\bibinfo {volume} {96}},\ \bibinfo
  {pages} {085418} (\bibinfo {year} {2017})}\BibitemShut {NoStop}%
\bibitem [{\citenamefont {Clarke}(2017)}]{Clarke2017}%
  \BibitemOpen
  \bibfield  {author} {\bibinfo {author} {\bibfnamefont {D.~J.}\ \bibnamefont
  {Clarke}},\ }\href {\doibase 10.1103/PhysRevB.96.201109} {\bibfield
  {journal} {\bibinfo  {journal} {Phys. Rev. B}\ }\textbf {\bibinfo {volume}
  {96}},\ \bibinfo {pages} {201109} (\bibinfo {year} {2017})}\BibitemShut
  {NoStop}%
\bibitem [{\citenamefont {Dourado}\ \emph {et~al.}(2024)\citenamefont
  {Dourado}, \citenamefont {Penteado},\ and\ \citenamefont
  {Egues}}]{Dourado2024}%
  \BibitemOpen
  \bibfield  {author} {\bibinfo {author} {\bibfnamefont {R.~A.}\ \bibnamefont
  {Dourado}}, \bibinfo {author} {\bibfnamefont {P.~H.}\ \bibnamefont
  {Penteado}}, \ and\ \bibinfo {author} {\bibfnamefont {J.~C.}\ \bibnamefont
  {Egues}},\ }\href {\doibase 10.1103/PhysRevB.110.014504} {\bibfield
  {journal} {\bibinfo  {journal} {Phys. Rev. B}\ }\textbf {\bibinfo {volume}
  {110}},\ \bibinfo {pages} {014504} (\bibinfo {year} {2024})}\BibitemShut
  {NoStop}%
\bibitem [{\citenamefont {Kleinherbers}\ \emph {et~al.}(2023)\citenamefont
  {Kleinherbers}, \citenamefont {Sch\"unemann},\ and\ \citenamefont
  {K\"onig}}]{Kleinherbers2023}%
  \BibitemOpen
  \bibfield  {author} {\bibinfo {author} {\bibfnamefont {E.}~\bibnamefont
  {Kleinherbers}}, \bibinfo {author} {\bibfnamefont {A.}~\bibnamefont
  {Sch\"unemann}}, \ and\ \bibinfo {author} {\bibfnamefont {J.}~\bibnamefont
  {K\"onig}},\ }\href {\doibase 10.1103/PhysRevB.107.195407} {\bibfield
  {journal} {\bibinfo  {journal} {Phys. Rev. B}\ }\textbf {\bibinfo {volume}
  {107}},\ \bibinfo {pages} {195407} (\bibinfo {year} {2023})}\BibitemShut
  {NoStop}%
\bibitem [{\citenamefont {Liu}\ \emph {et~al.}(2015{\natexlab{a}})\citenamefont
  {Liu}, \citenamefont {Cheng},\ and\ \citenamefont {Lutchyn}}]{Liu2015a}%
  \BibitemOpen
  \bibfield  {author} {\bibinfo {author} {\bibfnamefont {D.~E.}\ \bibnamefont
  {Liu}}, \bibinfo {author} {\bibfnamefont {M.}~\bibnamefont {Cheng}}, \ and\
  \bibinfo {author} {\bibfnamefont {R.~M.}\ \bibnamefont {Lutchyn}},\ }\href
  {\doibase 10.1103/PhysRevB.91.081405} {\bibfield  {journal} {\bibinfo
  {journal} {Phys. Rev. B}\ }\textbf {\bibinfo {volume} {91}},\ \bibinfo
  {pages} {081405} (\bibinfo {year} {2015}{\natexlab{a}})}\BibitemShut
  {NoStop}%
\bibitem [{\citenamefont {Liu}\ \emph {et~al.}(2015{\natexlab{b}})\citenamefont
  {Liu}, \citenamefont {Levchenko},\ and\ \citenamefont {Lutchyn}}]{Liu2015b}%
  \BibitemOpen
  \bibfield  {author} {\bibinfo {author} {\bibfnamefont {D.~E.}\ \bibnamefont
  {Liu}}, \bibinfo {author} {\bibfnamefont {A.}~\bibnamefont {Levchenko}}, \
  and\ \bibinfo {author} {\bibfnamefont {R.~M.}\ \bibnamefont {Lutchyn}},\
  }\href {\doibase 10.1103/PhysRevB.92.205422} {\bibfield  {journal} {\bibinfo
  {journal} {Phys. Rev. B}\ }\textbf {\bibinfo {volume} {92}},\ \bibinfo
  {pages} {205422} (\bibinfo {year} {2015}{\natexlab{b}})}\BibitemShut
  {NoStop}%
\bibitem [{\citenamefont {Smirnov}(2022)}]{Smirnov2022}%
  \BibitemOpen
  \bibfield  {author} {\bibinfo {author} {\bibfnamefont {S.}~\bibnamefont
  {Smirnov}},\ }\href {\doibase 10.1103/PhysRevB.105.205430} {\bibfield
  {journal} {\bibinfo  {journal} {Phys. Rev. B}\ }\textbf {\bibinfo {volume}
  {105}},\ \bibinfo {pages} {205430} (\bibinfo {year} {2022})}\BibitemShut
  {NoStop}%
\bibitem [{\citenamefont {Smirnov}(2024)}]{Smirnov2024}%
  \BibitemOpen
  \bibfield  {author} {\bibinfo {author} {\bibfnamefont {S.}~\bibnamefont
  {Smirnov}},\ }\href {\doibase 10.1103/PhysRevB.109.195410} {\bibfield
  {journal} {\bibinfo  {journal} {Phys. Rev. B}\ }\textbf {\bibinfo {volume}
  {109}},\ \bibinfo {pages} {195410} (\bibinfo {year} {2024})}\BibitemShut
  {NoStop}%
\bibitem [{\citenamefont {Mourik}\ \emph {et~al.}(2012)\citenamefont {Mourik},
  \citenamefont {Zuo}, \citenamefont {Frolov}, \citenamefont {Plissard},
  \citenamefont {Bakkers},\ and\ \citenamefont {Kouwenhoven}}]{Mourik2012}%
  \BibitemOpen
  \bibfield  {author} {\bibinfo {author} {\bibfnamefont {V.}~\bibnamefont
  {Mourik}}, \bibinfo {author} {\bibfnamefont {K.}~\bibnamefont {Zuo}},
  \bibinfo {author} {\bibfnamefont {S.~M.}\ \bibnamefont {Frolov}}, \bibinfo
  {author} {\bibfnamefont {S.~R.}\ \bibnamefont {Plissard}}, \bibinfo {author}
  {\bibfnamefont {E.~P. A.~M.}\ \bibnamefont {Bakkers}}, \ and\ \bibinfo
  {author} {\bibfnamefont {L.~P.}\ \bibnamefont {Kouwenhoven}},\ }\href
  {\doibase 10.1126/science.1222360} {\bibfield  {journal} {\bibinfo  {journal}
  {Science}\ }\textbf {\bibinfo {volume} {336}},\ \bibinfo {pages} {1003}
  (\bibinfo {year} {2012})},\ \Eprint
  {http://arxiv.org/abs/https://www.science.org/doi/pdf/10.1126/science.1222360}
  {https://www.science.org/doi/pdf/10.1126/science.1222360} \BibitemShut
  {NoStop}%
\bibitem [{\citenamefont {Deng}\ \emph {et~al.}(2012)\citenamefont {Deng},
  \citenamefont {Yu}, \citenamefont {Huang}, \citenamefont {Larsson},
  \citenamefont {Caroff},\ and\ \citenamefont {Xu}}]{Deng2012}%
  \BibitemOpen
  \bibfield  {author} {\bibinfo {author} {\bibfnamefont {M.~T.}\ \bibnamefont
  {Deng}}, \bibinfo {author} {\bibfnamefont {C.~L.}\ \bibnamefont {Yu}},
  \bibinfo {author} {\bibfnamefont {G.~Y.}\ \bibnamefont {Huang}}, \bibinfo
  {author} {\bibfnamefont {M.}~\bibnamefont {Larsson}}, \bibinfo {author}
  {\bibfnamefont {P.}~\bibnamefont {Caroff}}, \ and\ \bibinfo {author}
  {\bibfnamefont {H.~Q.}\ \bibnamefont {Xu}},\ }\href {\doibase
  10.1021/nl303758w} {\bibfield  {journal} {\bibinfo  {journal} {Nano Lett.}\
  }\textbf {\bibinfo {volume} {12}},\ \bibinfo {pages} {6414} (\bibinfo {year}
  {2012})}\BibitemShut {NoStop}%
\bibitem [{\citenamefont {Lee}\ \emph {et~al.}(2013)\citenamefont {Lee},
  \citenamefont {Jiang}, \citenamefont {Houzet}, \citenamefont {Aguado},
  \citenamefont {Lieber},\ and\ \citenamefont {Franceschi}}]{Lee2013}%
  \BibitemOpen
  \bibfield  {author} {\bibinfo {author} {\bibfnamefont {E.~J.~H.}\
  \bibnamefont {Lee}}, \bibinfo {author} {\bibfnamefont {X.}~\bibnamefont
  {Jiang}}, \bibinfo {author} {\bibfnamefont {M.}~\bibnamefont {Houzet}},
  \bibinfo {author} {\bibfnamefont {R.}~\bibnamefont {Aguado}}, \bibinfo
  {author} {\bibfnamefont {C.~M.}\ \bibnamefont {Lieber}}, \ and\ \bibinfo
  {author} {\bibfnamefont {S.~D.}\ \bibnamefont {Franceschi}},\ }\href
  {\doibase 10.1038/nnano.2013.267} {\bibfield  {journal} {\bibinfo  {journal}
  {Nat. Nanotechnol.}\ }\textbf {\bibinfo {volume} {9}},\ \bibinfo {pages} {79}
  (\bibinfo {year} {2013})}\BibitemShut {NoStop}%
\bibitem [{\citenamefont {Deng}\ \emph {et~al.}(2016)\citenamefont {Deng},
  \citenamefont {Vaitiekėnas}, \citenamefont {Hansen}, \citenamefont {Danon},
  \citenamefont {Leijnse}, \citenamefont {Flensberg}, \citenamefont {Nygård},
  \citenamefont {Krogstrup},\ and\ \citenamefont {Marcus}}]{Deng2016}%
  \BibitemOpen
  \bibfield  {author} {\bibinfo {author} {\bibfnamefont {M.~T.}\ \bibnamefont
  {Deng}}, \bibinfo {author} {\bibfnamefont {S.}~\bibnamefont {Vaitiekėnas}},
  \bibinfo {author} {\bibfnamefont {E.~B.}\ \bibnamefont {Hansen}}, \bibinfo
  {author} {\bibfnamefont {J.}~\bibnamefont {Danon}}, \bibinfo {author}
  {\bibfnamefont {M.}~\bibnamefont {Leijnse}}, \bibinfo {author} {\bibfnamefont
  {K.}~\bibnamefont {Flensberg}}, \bibinfo {author} {\bibfnamefont
  {J.}~\bibnamefont {Nygård}}, \bibinfo {author} {\bibfnamefont
  {P.}~\bibnamefont {Krogstrup}}, \ and\ \bibinfo {author} {\bibfnamefont
  {C.~M.}\ \bibnamefont {Marcus}},\ }\href {\doibase 10.1126/science.aaf3961}
  {\bibfield  {journal} {\bibinfo  {journal} {Science}\ }\textbf {\bibinfo
  {volume} {354}},\ \bibinfo {pages} {1557} (\bibinfo {year} {2016})},\ \Eprint
  {http://arxiv.org/abs/https://www.science.org/doi/pdf/10.1126/science.aaf3961}
  {https://www.science.org/doi/pdf/10.1126/science.aaf3961} \BibitemShut
  {NoStop}%
\bibitem [{\citenamefont {Aghaee}\ \emph {et~al.}(2023)\citenamefont {Aghaee},
  \citenamefont {Akkala}, \citenamefont {Alam}, \citenamefont {Ali},
  \citenamefont {Alcaraz~Ramirez}, \citenamefont {Andrzejczuk}, \citenamefont
  {Antipov}, \citenamefont {Aseev}, \citenamefont {Astafev}, \citenamefont
  {Bauer}, \citenamefont {Becker}, \citenamefont {Boddapati}, \citenamefont
  {Boekhout}, \citenamefont {Bommer}, \citenamefont {Bosma}, \citenamefont
  {Bourdet}, \citenamefont {Boutin}, \citenamefont {Caroff}, \citenamefont
  {Casparis}, \citenamefont {Cassidy}, \citenamefont {Chatoor}, \citenamefont
  {Christensen}, \citenamefont {Clay}, \citenamefont {Cole}, \citenamefont
  {Corsetti}, \citenamefont {Cui}, \citenamefont {Dalampiras}, \citenamefont
  {Dokania}, \citenamefont {de~Lange}, \citenamefont {de~Moor}, \citenamefont
  {Estrada Salda\~na}, \citenamefont {Fallahi}, \citenamefont {Fathabad},
  \citenamefont {Gamble}, \citenamefont {Gardner}, \citenamefont {Govender},
  \citenamefont {Griggio}, \citenamefont {Grigoryan}, \citenamefont {Gronin},
  \citenamefont {Gukelberger}, \citenamefont {Hansen}, \citenamefont {Heedt},
  \citenamefont {Herranz~Zamorano}, \citenamefont {Ho}, \citenamefont
  {Holgaard}, \citenamefont {Ingerslev}, \citenamefont {Johansson},
  \citenamefont {Jones}, \citenamefont {Kallaher}, \citenamefont {Karimi},
  \citenamefont {Karzig}, \citenamefont {King}, \citenamefont {Kloster},
  \citenamefont {Knapp}, \citenamefont {Kocon}, \citenamefont {Koski},
  \citenamefont {Kostamo}, \citenamefont {Krogstrup}, \citenamefont {Kumar},
  \citenamefont {Laeven}, \citenamefont {Larsen}, \citenamefont {Li},
  \citenamefont {Lindemann}, \citenamefont {Love}, \citenamefont {Lutchyn},
  \citenamefont {Madsen}, \citenamefont {Manfra}, \citenamefont {Markussen},
  \citenamefont {Martinez}, \citenamefont {McNeil}, \citenamefont {Memisevic},
  \citenamefont {Morgan}, \citenamefont {Mullally}, \citenamefont {Nayak},
  \citenamefont {Nielsen}, \citenamefont {Nielsen}, \citenamefont {Nijholt},
  \citenamefont {Nurmohamed}, \citenamefont {O'Farrell}, \citenamefont {Otani},
  \citenamefont {Pauka}, \citenamefont {Petersson}, \citenamefont {Petit},
  \citenamefont {Pikulin}, \citenamefont {Preiss}, \citenamefont
  {Quintero-Perez}, \citenamefont {Rajpalke}, \citenamefont {Rasmussen},
  \citenamefont {Razmadze}, \citenamefont {Reentila}, \citenamefont {Reilly},
  \citenamefont {Rouse}, \citenamefont {Sadovskyy}, \citenamefont {Sainiemi},
  \citenamefont {Schreppler}, \citenamefont {Sidorkin}, \citenamefont {Singh},
  \citenamefont {Singh}, \citenamefont {Sinha}, \citenamefont {Sohr},
  \citenamefont {Stankevi\ifmmode~\check{c}\else \v{c}\fi{}}, \citenamefont
  {Stek}, \citenamefont {Suominen}, \citenamefont {Suter}, \citenamefont
  {Svidenko}, \citenamefont {Teicher}, \citenamefont {Temuerhan}, \citenamefont
  {Thiyagarajah}, \citenamefont {Tholapi}, \citenamefont {Thomas},
  \citenamefont {Toomey}, \citenamefont {Upadhyay}, \citenamefont {Urban},
  \citenamefont {Vaitiek\ifmmode~\dot{e}\else \.{e}\fi{}nas}, \citenamefont
  {Van~Hoogdalem}, \citenamefont {Van~Woerkom}, \citenamefont {Viazmitinov},
  \citenamefont {Vogel}, \citenamefont {Waddy}, \citenamefont {Watson},
  \citenamefont {Weston}, \citenamefont {Winkler}, \citenamefont {Yang},
  \citenamefont {Yau}, \citenamefont {Yi}, \citenamefont {Yucelen},
  \citenamefont {Webster}, \citenamefont {Zeisel},\ and\ \citenamefont
  {Zhao}}]{Aghaee2023}%
  \BibitemOpen
  \bibfield  {author} {\bibinfo {author} {\bibfnamefont {M.}~\bibnamefont
  {Aghaee}}, \bibinfo {author} {\bibfnamefont {A.}~\bibnamefont {Akkala}},
  \bibinfo {author} {\bibfnamefont {Z.}~\bibnamefont {Alam}}, \bibinfo {author}
  {\bibfnamefont {R.}~\bibnamefont {Ali}}, \bibinfo {author} {\bibfnamefont
  {A.}~\bibnamefont {Alcaraz~Ramirez}}, \bibinfo {author} {\bibfnamefont
  {M.}~\bibnamefont {Andrzejczuk}}, \bibinfo {author} {\bibfnamefont {A.~E.}\
  \bibnamefont {Antipov}}, \bibinfo {author} {\bibfnamefont {P.}~\bibnamefont
  {Aseev}}, \bibinfo {author} {\bibfnamefont {M.}~\bibnamefont {Astafev}},
  \bibinfo {author} {\bibfnamefont {B.}~\bibnamefont {Bauer}}, \bibinfo
  {author} {\bibfnamefont {J.}~\bibnamefont {Becker}}, \bibinfo {author}
  {\bibfnamefont {S.}~\bibnamefont {Boddapati}}, \bibinfo {author}
  {\bibfnamefont {F.}~\bibnamefont {Boekhout}}, \bibinfo {author}
  {\bibfnamefont {J.}~\bibnamefont {Bommer}}, \bibinfo {author} {\bibfnamefont
  {T.}~\bibnamefont {Bosma}}, \bibinfo {author} {\bibfnamefont
  {L.}~\bibnamefont {Bourdet}}, \bibinfo {author} {\bibfnamefont
  {S.}~\bibnamefont {Boutin}}, \bibinfo {author} {\bibfnamefont
  {P.}~\bibnamefont {Caroff}}, \bibinfo {author} {\bibfnamefont
  {L.}~\bibnamefont {Casparis}}, \bibinfo {author} {\bibfnamefont
  {M.}~\bibnamefont {Cassidy}}, \bibinfo {author} {\bibfnamefont
  {S.}~\bibnamefont {Chatoor}}, \bibinfo {author} {\bibfnamefont {A.~W.}\
  \bibnamefont {Christensen}}, \bibinfo {author} {\bibfnamefont
  {N.}~\bibnamefont {Clay}}, \bibinfo {author} {\bibfnamefont {W.~S.}\
  \bibnamefont {Cole}}, \bibinfo {author} {\bibfnamefont {F.}~\bibnamefont
  {Corsetti}}, \bibinfo {author} {\bibfnamefont {A.}~\bibnamefont {Cui}},
  \bibinfo {author} {\bibfnamefont {P.}~\bibnamefont {Dalampiras}}, \bibinfo
  {author} {\bibfnamefont {A.}~\bibnamefont {Dokania}}, \bibinfo {author}
  {\bibfnamefont {G.}~\bibnamefont {de~Lange}}, \bibinfo {author}
  {\bibfnamefont {M.}~\bibnamefont {de~Moor}}, \bibinfo {author} {\bibfnamefont
  {J.~C.}\ \bibnamefont {Estrada Salda\~na}}, \bibinfo {author} {\bibfnamefont
  {S.}~\bibnamefont {Fallahi}}, \bibinfo {author} {\bibfnamefont {Z.~H.}\
  \bibnamefont {Fathabad}}, \bibinfo {author} {\bibfnamefont {J.}~\bibnamefont
  {Gamble}}, \bibinfo {author} {\bibfnamefont {G.}~\bibnamefont {Gardner}},
  \bibinfo {author} {\bibfnamefont {D.}~\bibnamefont {Govender}}, \bibinfo
  {author} {\bibfnamefont {F.}~\bibnamefont {Griggio}}, \bibinfo {author}
  {\bibfnamefont {R.}~\bibnamefont {Grigoryan}}, \bibinfo {author}
  {\bibfnamefont {S.}~\bibnamefont {Gronin}}, \bibinfo {author} {\bibfnamefont
  {J.}~\bibnamefont {Gukelberger}}, \bibinfo {author} {\bibfnamefont {E.~B.}\
  \bibnamefont {Hansen}}, \bibinfo {author} {\bibfnamefont {S.}~\bibnamefont
  {Heedt}}, \bibinfo {author} {\bibfnamefont {J.}~\bibnamefont
  {Herranz~Zamorano}}, \bibinfo {author} {\bibfnamefont {S.}~\bibnamefont
  {Ho}}, \bibinfo {author} {\bibfnamefont {U.~L.}\ \bibnamefont {Holgaard}},
  \bibinfo {author} {\bibfnamefont {H.}~\bibnamefont {Ingerslev}}, \bibinfo
  {author} {\bibfnamefont {L.}~\bibnamefont {Johansson}}, \bibinfo {author}
  {\bibfnamefont {J.}~\bibnamefont {Jones}}, \bibinfo {author} {\bibfnamefont
  {R.}~\bibnamefont {Kallaher}}, \bibinfo {author} {\bibfnamefont
  {F.}~\bibnamefont {Karimi}}, \bibinfo {author} {\bibfnamefont
  {T.}~\bibnamefont {Karzig}}, \bibinfo {author} {\bibfnamefont
  {E.}~\bibnamefont {King}}, \bibinfo {author} {\bibfnamefont {M.~E.}\
  \bibnamefont {Kloster}}, \bibinfo {author} {\bibfnamefont {C.}~\bibnamefont
  {Knapp}}, \bibinfo {author} {\bibfnamefont {D.}~\bibnamefont {Kocon}},
  \bibinfo {author} {\bibfnamefont {J.}~\bibnamefont {Koski}}, \bibinfo
  {author} {\bibfnamefont {P.}~\bibnamefont {Kostamo}}, \bibinfo {author}
  {\bibfnamefont {P.}~\bibnamefont {Krogstrup}}, \bibinfo {author}
  {\bibfnamefont {M.}~\bibnamefont {Kumar}}, \bibinfo {author} {\bibfnamefont
  {T.}~\bibnamefont {Laeven}}, \bibinfo {author} {\bibfnamefont
  {T.}~\bibnamefont {Larsen}}, \bibinfo {author} {\bibfnamefont
  {K.}~\bibnamefont {Li}}, \bibinfo {author} {\bibfnamefont {T.}~\bibnamefont
  {Lindemann}}, \bibinfo {author} {\bibfnamefont {J.}~\bibnamefont {Love}},
  \bibinfo {author} {\bibfnamefont {R.}~\bibnamefont {Lutchyn}}, \bibinfo
  {author} {\bibfnamefont {M.~H.}\ \bibnamefont {Madsen}}, \bibinfo {author}
  {\bibfnamefont {M.}~\bibnamefont {Manfra}}, \bibinfo {author} {\bibfnamefont
  {S.}~\bibnamefont {Markussen}}, \bibinfo {author} {\bibfnamefont
  {E.}~\bibnamefont {Martinez}}, \bibinfo {author} {\bibfnamefont
  {R.}~\bibnamefont {McNeil}}, \bibinfo {author} {\bibfnamefont
  {E.}~\bibnamefont {Memisevic}}, \bibinfo {author} {\bibfnamefont
  {T.}~\bibnamefont {Morgan}}, \bibinfo {author} {\bibfnamefont
  {A.}~\bibnamefont {Mullally}}, \bibinfo {author} {\bibfnamefont
  {C.}~\bibnamefont {Nayak}}, \bibinfo {author} {\bibfnamefont
  {J.}~\bibnamefont {Nielsen}}, \bibinfo {author} {\bibfnamefont {W.~H.~P.}\
  \bibnamefont {Nielsen}}, \bibinfo {author} {\bibfnamefont {B.}~\bibnamefont
  {Nijholt}}, \bibinfo {author} {\bibfnamefont {A.}~\bibnamefont {Nurmohamed}},
  \bibinfo {author} {\bibfnamefont {E.}~\bibnamefont {O'Farrell}}, \bibinfo
  {author} {\bibfnamefont {K.}~\bibnamefont {Otani}}, \bibinfo {author}
  {\bibfnamefont {S.}~\bibnamefont {Pauka}}, \bibinfo {author} {\bibfnamefont
  {K.}~\bibnamefont {Petersson}}, \bibinfo {author} {\bibfnamefont
  {L.}~\bibnamefont {Petit}}, \bibinfo {author} {\bibfnamefont {D.~I.}\
  \bibnamefont {Pikulin}}, \bibinfo {author} {\bibfnamefont {F.}~\bibnamefont
  {Preiss}}, \bibinfo {author} {\bibfnamefont {M.}~\bibnamefont
  {Quintero-Perez}}, \bibinfo {author} {\bibfnamefont {M.}~\bibnamefont
  {Rajpalke}}, \bibinfo {author} {\bibfnamefont {K.}~\bibnamefont {Rasmussen}},
  \bibinfo {author} {\bibfnamefont {D.}~\bibnamefont {Razmadze}}, \bibinfo
  {author} {\bibfnamefont {O.}~\bibnamefont {Reentila}}, \bibinfo {author}
  {\bibfnamefont {D.}~\bibnamefont {Reilly}}, \bibinfo {author} {\bibfnamefont
  {R.}~\bibnamefont {Rouse}}, \bibinfo {author} {\bibfnamefont
  {I.}~\bibnamefont {Sadovskyy}}, \bibinfo {author} {\bibfnamefont
  {L.}~\bibnamefont {Sainiemi}}, \bibinfo {author} {\bibfnamefont
  {S.}~\bibnamefont {Schreppler}}, \bibinfo {author} {\bibfnamefont
  {V.}~\bibnamefont {Sidorkin}}, \bibinfo {author} {\bibfnamefont
  {A.}~\bibnamefont {Singh}}, \bibinfo {author} {\bibfnamefont
  {S.}~\bibnamefont {Singh}}, \bibinfo {author} {\bibfnamefont
  {S.}~\bibnamefont {Sinha}}, \bibinfo {author} {\bibfnamefont
  {P.}~\bibnamefont {Sohr}}, \bibinfo {author} {\bibfnamefont {T.~c.~v.}\
  \bibnamefont {Stankevi\ifmmode~\check{c}\else \v{c}\fi{}}}, \bibinfo {author}
  {\bibfnamefont {L.}~\bibnamefont {Stek}}, \bibinfo {author} {\bibfnamefont
  {H.}~\bibnamefont {Suominen}}, \bibinfo {author} {\bibfnamefont
  {J.}~\bibnamefont {Suter}}, \bibinfo {author} {\bibfnamefont
  {V.}~\bibnamefont {Svidenko}}, \bibinfo {author} {\bibfnamefont
  {S.}~\bibnamefont {Teicher}}, \bibinfo {author} {\bibfnamefont
  {M.}~\bibnamefont {Temuerhan}}, \bibinfo {author} {\bibfnamefont
  {N.}~\bibnamefont {Thiyagarajah}}, \bibinfo {author} {\bibfnamefont
  {R.}~\bibnamefont {Tholapi}}, \bibinfo {author} {\bibfnamefont
  {M.}~\bibnamefont {Thomas}}, \bibinfo {author} {\bibfnamefont
  {E.}~\bibnamefont {Toomey}}, \bibinfo {author} {\bibfnamefont
  {S.}~\bibnamefont {Upadhyay}}, \bibinfo {author} {\bibfnamefont
  {I.}~\bibnamefont {Urban}}, \bibinfo {author} {\bibfnamefont
  {S.}~\bibnamefont {Vaitiek\ifmmode~\dot{e}\else \.{e}\fi{}nas}}, \bibinfo
  {author} {\bibfnamefont {K.}~\bibnamefont {Van~Hoogdalem}}, \bibinfo {author}
  {\bibfnamefont {D.}~\bibnamefont {Van~Woerkom}}, \bibinfo {author}
  {\bibfnamefont {D.~V.}\ \bibnamefont {Viazmitinov}}, \bibinfo {author}
  {\bibfnamefont {D.}~\bibnamefont {Vogel}}, \bibinfo {author} {\bibfnamefont
  {S.}~\bibnamefont {Waddy}}, \bibinfo {author} {\bibfnamefont
  {J.}~\bibnamefont {Watson}}, \bibinfo {author} {\bibfnamefont
  {J.}~\bibnamefont {Weston}}, \bibinfo {author} {\bibfnamefont {G.~W.}\
  \bibnamefont {Winkler}}, \bibinfo {author} {\bibfnamefont {C.~K.}\
  \bibnamefont {Yang}}, \bibinfo {author} {\bibfnamefont {S.}~\bibnamefont
  {Yau}}, \bibinfo {author} {\bibfnamefont {D.}~\bibnamefont {Yi}}, \bibinfo
  {author} {\bibfnamefont {E.}~\bibnamefont {Yucelen}}, \bibinfo {author}
  {\bibfnamefont {A.}~\bibnamefont {Webster}}, \bibinfo {author} {\bibfnamefont
  {R.}~\bibnamefont {Zeisel}}, \ and\ \bibinfo {author} {\bibfnamefont
  {R.}~\bibnamefont {Zhao}} (\bibinfo {collaboration} {Microsoft Quantum}),\
  }\href {\doibase 10.1103/PhysRevB.107.245423} {\bibfield  {journal} {\bibinfo
   {journal} {Phys. Rev. B}\ }\textbf {\bibinfo {volume} {107}},\ \bibinfo
  {pages} {245423} (\bibinfo {year} {2023})}\BibitemShut {NoStop}%
\bibitem [{sup()}]{supplemental}%
  \BibitemOpen
  \href@noop {} {\enquote {\bibinfo {title} {See supplemental material at [url
  will be inserted by publisher] for details.}}\ }\BibitemShut {NoStop}%
\bibitem [{\citenamefont {Auerbach}(1994)}]{Auerbach1994}%
  \BibitemOpen
  \bibfield  {author} {\bibinfo {author} {\bibfnamefont {A.}~\bibnamefont
  {Auerbach}},\ }\href@noop {} {\emph {\bibinfo {title} {Interacting Electrons
  and Quantum Magnetism}}}\ (\bibinfo  {publisher} {Springer-Verlag},\ \bibinfo
  {year} {1994})\BibitemShut {NoStop}%
\bibitem [{\citenamefont {Blum}(1996)}]{Blum1996}%
  \BibitemOpen
  \bibfield  {author} {\bibinfo {author} {\bibfnamefont {K.}~\bibnamefont
  {Blum}},\ }\enquote {\bibinfo {title} {Quantum theory of relaxation},}\ in\
  \href {\doibase 10.1007/978-1-4757-4931-1_8} {\emph {\bibinfo {booktitle}
  {Density Matrix Theory and Applications}}}\ (\bibinfo  {publisher} {Springer
  US},\ \bibinfo {address} {Boston, MA},\ \bibinfo {year} {1996})\ pp.\
  \bibinfo {pages} {261--302}\BibitemShut {NoStop}%
\bibitem [{\citenamefont {Probst}\ \emph {et~al.}(2022)\citenamefont {Probst},
  \citenamefont {Virtanen},\ and\ \citenamefont {Recher}}]{Probst2022}%
  \BibitemOpen
  \bibfield  {author} {\bibinfo {author} {\bibfnamefont {B.}~\bibnamefont
  {Probst}}, \bibinfo {author} {\bibfnamefont {P.}~\bibnamefont {Virtanen}}, \
  and\ \bibinfo {author} {\bibfnamefont {P.}~\bibnamefont {Recher}},\ }\href
  {\doibase 10.1103/PhysRevB.106.085406} {\bibfield  {journal} {\bibinfo
  {journal} {Phys. Rev. B}\ }\textbf {\bibinfo {volume} {106}},\ \bibinfo
  {pages} {085406} (\bibinfo {year} {2022})}\BibitemShut {NoStop}%
\bibitem [{\citenamefont {Bagrets}\ and\ \citenamefont
  {Nazarov}(2003)}]{Bagrets2003}%
  \BibitemOpen
  \bibfield  {author} {\bibinfo {author} {\bibfnamefont {D.~A.}\ \bibnamefont
  {Bagrets}}\ and\ \bibinfo {author} {\bibfnamefont {Y.~V.}\ \bibnamefont
  {Nazarov}},\ }\href {\doibase 10.1103/PhysRevB.67.085316} {\bibfield
  {journal} {\bibinfo  {journal} {Phys. Rev. B}\ }\textbf {\bibinfo {volume}
  {67}},\ \bibinfo {pages} {085316} (\bibinfo {year} {2003})}\BibitemShut
  {NoStop}%
\bibitem [{\citenamefont {Flindt}\ \emph {et~al.}(2010)\citenamefont {Flindt},
  \citenamefont {{Novotn\'y}}, \citenamefont {Braggio},\ and\ \citenamefont
  {Jauho}}]{Flindt2010}%
  \BibitemOpen
  \bibfield  {author} {\bibinfo {author} {\bibfnamefont {C.}~\bibnamefont
  {Flindt}}, \bibinfo {author} {\bibfnamefont {T.}~\bibnamefont {{Novotn\'y}}},
  \bibinfo {author} {\bibfnamefont {A.}~\bibnamefont {Braggio}}, \ and\
  \bibinfo {author} {\bibfnamefont {A.-P.}\ \bibnamefont {Jauho}},\ }\href
  {\doibase 10.1103/PhysRevB.82.155407} {\bibfield  {journal} {\bibinfo
  {journal} {Phys. Rev. B}\ }\textbf {\bibinfo {volume} {82}},\ \bibinfo
  {pages} {155407} (\bibinfo {year} {2010})}\BibitemShut {NoStop}%
\bibitem [{\citenamefont {Fano}(1947)}]{Fano1947}%
  \BibitemOpen
  \bibfield  {author} {\bibinfo {author} {\bibfnamefont {U.}~\bibnamefont
  {Fano}},\ }\href {\doibase 10.1103/PhysRev.72.26} {\bibfield  {journal}
  {\bibinfo  {journal} {Phys. Rev.}\ }\textbf {\bibinfo {volume} {72}},\
  \bibinfo {pages} {26} (\bibinfo {year} {1947})}\BibitemShut {NoStop}%
\bibitem [{\citenamefont {Bittermann}\ \emph {et~al.}(2022)\citenamefont
  {Bittermann}, \citenamefont {De~Beule}, \citenamefont {Frombach},\ and\
  \citenamefont {Recher}}]{Bittermann2022}%
  \BibitemOpen
  \bibfield  {author} {\bibinfo {author} {\bibfnamefont {L.}~\bibnamefont
  {Bittermann}}, \bibinfo {author} {\bibfnamefont {C.}~\bibnamefont
  {De~Beule}}, \bibinfo {author} {\bibfnamefont {D.}~\bibnamefont {Frombach}},
  \ and\ \bibinfo {author} {\bibfnamefont {P.}~\bibnamefont {Recher}},\ }\href
  {\doibase 10.1103/PhysRevB.106.075305} {\bibfield  {journal} {\bibinfo
  {journal} {Phys. Rev. B}\ }\textbf {\bibinfo {volume} {106}},\ \bibinfo
  {pages} {075305} (\bibinfo {year} {2022})}\BibitemShut {NoStop}%
\bibitem [{\citenamefont {Pikulin}\ and\ \citenamefont
  {Nazarov}(2012)}]{Pikulin2012}%
  \BibitemOpen
  \bibfield  {author} {\bibinfo {author} {\bibfnamefont {D.~I.}\ \bibnamefont
  {Pikulin}}\ and\ \bibinfo {author} {\bibfnamefont {Y.~V.}\ \bibnamefont
  {Nazarov}},\ }\href {\doibase 10.1103/PhysRevB.86.140504} {\bibfield
  {journal} {\bibinfo  {journal} {Phys. Rev. B}\ }\textbf {\bibinfo {volume}
  {86}},\ \bibinfo {pages} {140504} (\bibinfo {year} {2012})}\BibitemShut
  {NoStop}%
\bibitem [{\citenamefont {Cheng}\ \emph {et~al.}(2012)\citenamefont {Cheng},
  \citenamefont {Lutchyn},\ and\ \citenamefont {Das~Sarma}}]{Cheng2012}%
  \BibitemOpen
  \bibfield  {author} {\bibinfo {author} {\bibfnamefont {M.}~\bibnamefont
  {Cheng}}, \bibinfo {author} {\bibfnamefont {R.~M.}\ \bibnamefont {Lutchyn}},
  \ and\ \bibinfo {author} {\bibfnamefont {S.}~\bibnamefont {Das~Sarma}},\
  }\href {\doibase 10.1103/PhysRevB.85.165124} {\bibfield  {journal} {\bibinfo
  {journal} {Phys. Rev. B}\ }\textbf {\bibinfo {volume} {85}},\ \bibinfo
  {pages} {165124} (\bibinfo {year} {2012})}\BibitemShut {NoStop}%
\bibitem [{\citenamefont {San-Jose}\ \emph {et~al.}(2012)\citenamefont
  {San-Jose}, \citenamefont {Prada},\ and\ \citenamefont
  {Aguado}}]{SanJose2012}%
  \BibitemOpen
  \bibfield  {author} {\bibinfo {author} {\bibfnamefont {P.}~\bibnamefont
  {San-Jose}}, \bibinfo {author} {\bibfnamefont {E.}~\bibnamefont {Prada}}, \
  and\ \bibinfo {author} {\bibfnamefont {R.}~\bibnamefont {Aguado}},\ }\href
  {\doibase 10.1103/PhysRevLett.108.257001} {\bibfield  {journal} {\bibinfo
  {journal} {Phys. Rev. Lett.}\ }\textbf {\bibinfo {volume} {108}},\ \bibinfo
  {pages} {257001} (\bibinfo {year} {2012})}\BibitemShut {NoStop}%
\bibitem [{\citenamefont {Virtanen}\ and\ \citenamefont
  {Recher}(2013)}]{Virtanen2013}%
  \BibitemOpen
  \bibfield  {author} {\bibinfo {author} {\bibfnamefont {P.}~\bibnamefont
  {Virtanen}}\ and\ \bibinfo {author} {\bibfnamefont {P.}~\bibnamefont
  {Recher}},\ }\href {\doibase 10.1103/PhysRevB.88.144507} {\bibfield
  {journal} {\bibinfo  {journal} {Phys. Rev. B}\ }\textbf {\bibinfo {volume}
  {88}},\ \bibinfo {pages} {144507} (\bibinfo {year} {2013})}\BibitemShut
  {NoStop}%
\bibitem [{Note1()}]{Note1}%
  \BibitemOpen
  \bibinfo {note} {The absence of cross correlations $S_{LR}^{ST}=0$ in the
  ST-only result is correct for any bias profile and a consequence of $\partial
  _{\chi _i}W_{10}|_{\chi = 0} = \partial _{\chi _i}W_{01}|_{\chi = 0} $ for $i
  = L,R$, which is always the case when the anomalous and normal tunneling
  events have equal weights, hence we regard this property as a
  Majorana-specific result.}\BibitemShut {Stop}%
\bibitem [{\citenamefont {Colbert}\ and\ \citenamefont
  {Lee}(2014)}]{Colbert2014}%
  \BibitemOpen
  \bibfield  {author} {\bibinfo {author} {\bibfnamefont {J.~R.}\ \bibnamefont
  {Colbert}}\ and\ \bibinfo {author} {\bibfnamefont {P.~A.}\ \bibnamefont
  {Lee}},\ }\href {\doibase 10.1103/PhysRevB.89.140505} {\bibfield  {journal}
  {\bibinfo  {journal} {Phys. Rev. B}\ }\textbf {\bibinfo {volume} {89}},\
  \bibinfo {pages} {140505} (\bibinfo {year} {2014})}\BibitemShut {NoStop}%
\bibitem [{\citenamefont {Jonckheere}\ \emph {et~al.}(2019)\citenamefont
  {Jonckheere}, \citenamefont {Rech}, \citenamefont {Zazunov}, \citenamefont
  {Egger}, \citenamefont {Yeyati},\ and\ \citenamefont
  {Martin}}]{Jonckheere2019}%
  \BibitemOpen
  \bibfield  {author} {\bibinfo {author} {\bibfnamefont {T.}~\bibnamefont
  {Jonckheere}}, \bibinfo {author} {\bibfnamefont {J.}~\bibnamefont {Rech}},
  \bibinfo {author} {\bibfnamefont {A.}~\bibnamefont {Zazunov}}, \bibinfo
  {author} {\bibfnamefont {R.}~\bibnamefont {Egger}}, \bibinfo {author}
  {\bibfnamefont {A.~L.}\ \bibnamefont {Yeyati}}, \ and\ \bibinfo {author}
  {\bibfnamefont {T.}~\bibnamefont {Martin}},\ }\href {\doibase
  10.1103/PhysRevLett.122.097003} {\bibfield  {journal} {\bibinfo  {journal}
  {Phys. Rev. Lett.}\ }\textbf {\bibinfo {volume} {122}},\ \bibinfo {pages}
  {097003} (\bibinfo {year} {2019})}\BibitemShut {NoStop}%
\end{thebibliography}%


\begin{thebibliography}{7}%
\makeatletter
\providecommand \@ifxundefined [1]{%
 \@ifx{#1\undefined}
}%
\providecommand \@ifnum [1]{%
 \ifnum #1\expandafter \@firstoftwo
 \else \expandafter \@secondoftwo
 \fi
}%
\providecommand \@ifx [1]{%
 \ifx #1\expandafter \@firstoftwo
 \else \expandafter \@secondoftwo
 \fi
}%
\providecommand \natexlab [1]{#1}%
\providecommand \enquote  [1]{``#1''}%
\providecommand \bibnamefont  [1]{#1}%
\providecommand \bibfnamefont [1]{#1}%
\providecommand \citenamefont [1]{#1}%
\providecommand \href@noop [0]{\@secondoftwo}%
\providecommand \href [0]{\begingroup \@sanitize@url \@href}%
\providecommand \@href[1]{\@@startlink{#1}\@@href}%
\providecommand \@@href[1]{\endgroup#1\@@endlink}%
\providecommand \@sanitize@url [0]{\catcode `\\12\catcode `\$12\catcode
  `\&12\catcode `\#12\catcode `\^12\catcode `\_12\catcode `\%12\relax}%
\providecommand \@@startlink[1]{}%
\providecommand \@@endlink[0]{}%
\providecommand \url  [0]{\begingroup\@sanitize@url \@url }%
\providecommand \@url [1]{\endgroup\@href {#1}{\urlprefix }}%
\providecommand \urlprefix  [0]{URL }%
\providecommand \Eprint [0]{\href }%
\providecommand \doibase [0]{http://dx.doi.org/}%
\providecommand \selectlanguage [0]{\@gobble}%
\providecommand \bibinfo  [0]{\@secondoftwo}%
\providecommand \bibfield  [0]{\@secondoftwo}%
\providecommand \translation [1]{[#1]}%
\providecommand \BibitemOpen [0]{}%
\providecommand \bibitemStop [0]{}%
\providecommand \bibitemNoStop [0]{.\EOS\space}%
\providecommand \EOS [0]{\spacefactor3000\relax}%
\providecommand \BibitemShut  [1]{\csname bibitem#1\endcsname}%
\let\auto@bib@innerbib\@empty
\bibitem [{\citenamefont {Auerbach}(1994)}]{Auerbach1994}%
  \BibitemOpen
  \bibfield  {author} {\bibinfo {author} {\bibfnamefont {A.}~\bibnamefont
  {Auerbach}},\ }\href@noop {} {\emph {\bibinfo {title} {Interacting Electrons
  and Quantum Magnetism}}}\ (\bibinfo  {publisher} {Springer-Verlag},\ \bibinfo
  {year} {1994})\BibitemShut {NoStop}%
\bibitem [{\citenamefont {Nilsson}\ \emph {et~al.}(2008)\citenamefont
  {Nilsson}, \citenamefont {Akhmerov},\ and\ \citenamefont
  {Beenakker}}]{Nilsson2008}%
  \BibitemOpen
  \bibfield  {author} {\bibinfo {author} {\bibfnamefont {J.}~\bibnamefont
  {Nilsson}}, \bibinfo {author} {\bibfnamefont {A.~R.}\ \bibnamefont
  {Akhmerov}}, \ and\ \bibinfo {author} {\bibfnamefont {C.~W.~J.}\ \bibnamefont
  {Beenakker}},\ }\href {\doibase 10.1103/PhysRevLett.101.120403} {\bibfield
  {journal} {\bibinfo  {journal} {Phys. Rev. Lett.}\ }\textbf {\bibinfo
  {volume} {101}},\ \bibinfo {pages} {120403} (\bibinfo {year}
  {2008})}\BibitemShut {NoStop}%
\bibitem [{\citenamefont {Koch}\ \emph {et~al.}(2004)\citenamefont {Koch},
  \citenamefont {von Oppen}, \citenamefont {Oreg},\ and\ \citenamefont
  {Sela}}]{Koch2004}%
  \BibitemOpen
  \bibfield  {author} {\bibinfo {author} {\bibfnamefont {J.}~\bibnamefont
  {Koch}}, \bibinfo {author} {\bibfnamefont {F.}~\bibnamefont {von Oppen}},
  \bibinfo {author} {\bibfnamefont {Y.}~\bibnamefont {Oreg}}, \ and\ \bibinfo
  {author} {\bibfnamefont {E.}~\bibnamefont {Sela}},\ }\href {\doibase
  10.1103/PhysRevB.70.195107} {\bibfield  {journal} {\bibinfo  {journal} {Phys.
  Rev. B}\ }\textbf {\bibinfo {volume} {70}},\ \bibinfo {pages} {195107}
  (\bibinfo {year} {2004})}\BibitemShut {NoStop}%
\bibitem [{\citenamefont {Koch}\ \emph {et~al.}(2006)\citenamefont {Koch},
  \citenamefont {von Oppen},\ and\ \citenamefont {Andreev}}]{Koch2006}%
  \BibitemOpen
  \bibfield  {author} {\bibinfo {author} {\bibfnamefont {J.}~\bibnamefont
  {Koch}}, \bibinfo {author} {\bibfnamefont {F.}~\bibnamefont {von Oppen}}, \
  and\ \bibinfo {author} {\bibfnamefont {A.~V.}\ \bibnamefont {Andreev}},\
  }\href {\doibase 10.1103/PhysRevB.74.205438} {\bibfield  {journal} {\bibinfo
  {journal} {Phys. Rev. B}\ }\textbf {\bibinfo {volume} {74}},\ \bibinfo
  {pages} {205438} (\bibinfo {year} {2006})}\BibitemShut {NoStop}%
\bibitem [{\citenamefont {Probst}\ \emph {et~al.}(2022)\citenamefont {Probst},
  \citenamefont {Virtanen},\ and\ \citenamefont {Recher}}]{Probst2022}%
  \BibitemOpen
  \bibfield  {author} {\bibinfo {author} {\bibfnamefont {B.}~\bibnamefont
  {Probst}}, \bibinfo {author} {\bibfnamefont {P.}~\bibnamefont {Virtanen}}, \
  and\ \bibinfo {author} {\bibfnamefont {P.}~\bibnamefont {Recher}},\ }\href
  {\doibase 10.1103/PhysRevB.106.085406} {\bibfield  {journal} {\bibinfo
  {journal} {Phys. Rev. B}\ }\textbf {\bibinfo {volume} {106}},\ \bibinfo
  {pages} {085406} (\bibinfo {year} {2022})}\BibitemShut {NoStop}%
\bibitem [{\citenamefont {Flindt}\ \emph {et~al.}(2010)\citenamefont {Flindt},
  \citenamefont {{Novotn\'y}}, \citenamefont {Braggio},\ and\ \citenamefont
  {Jauho}}]{Flindt2010}%
  \BibitemOpen
  \bibfield  {author} {\bibinfo {author} {\bibfnamefont {C.}~\bibnamefont
  {Flindt}}, \bibinfo {author} {\bibfnamefont {T.}~\bibnamefont {{Novotn\'y}}},
  \bibinfo {author} {\bibfnamefont {A.}~\bibnamefont {Braggio}}, \ and\
  \bibinfo {author} {\bibfnamefont {A.-P.}\ \bibnamefont {Jauho}},\ }\href
  {\doibase 10.1103/PhysRevB.82.155407} {\bibfield  {journal} {\bibinfo
  {journal} {Phys. Rev. B}\ }\textbf {\bibinfo {volume} {82}},\ \bibinfo
  {pages} {155407} (\bibinfo {year} {2010})}\BibitemShut {NoStop}%
\bibitem [{\citenamefont {Schuray}\ \emph {et~al.}(2017)\citenamefont
  {Schuray}, \citenamefont {Weithofer},\ and\ \citenamefont
  {Recher}}]{Schuray2017}%
  \BibitemOpen
  \bibfield  {author} {\bibinfo {author} {\bibfnamefont {A.}~\bibnamefont
  {Schuray}}, \bibinfo {author} {\bibfnamefont {L.}~\bibnamefont {Weithofer}},
  \ and\ \bibinfo {author} {\bibfnamefont {P.}~\bibnamefont {Recher}},\ }\href
  {\doibase 10.1103/PhysRevB.96.085417} {\bibfield  {journal} {\bibinfo
  {journal} {Phys. Rev. B}\ }\textbf {\bibinfo {volume} {96}},\ \bibinfo
  {pages} {085417} (\bibinfo {year} {2017})}\BibitemShut {NoStop}%
\end{thebibliography}%

\end{document}


\title{Supplemental material for: Super-Poissonian noise from quasiparticle poisoning in electron transport through a pair of Majorana bound states}
\author{Florinda Vi\~nas Bostr\"om}
\affiliation{Institut f\"ur Mathematische Physik, Technische Universit\"at Braunschweig, D-38106 Braunschweig, Germany}
\affiliation{Division of Solid State Physics and NanoLund, Lund University, Box 118, S-221 00 Lund, Sweden}
\author{Patrik Recher}
\affiliation{Institut f\"ur Mathematische Physik, Technische Universit\"at Braunschweig, D-38106 Braunschweig, Germany}
\affiliation{Laboratory of Emerging Nanometrology, D-38106 Braunschweig, Germany}
\date{\today} 


\maketitle


\tableofcontents

\section{Master equation and full counting statistics: sequential tunneling}

The sequential tunneling (ST) rates $W_{mn}$ in the Master equation (Eq.~(5) of the main text) are defined as
\begin{equation}
W_{mn}(\chi) = \Gamma^+_{nmmn}(\chi) + \Gamma^-_{nmmn}(\chi). \label{eq_Wmn}
\end{equation}
where 
\begin{equation}
\begin{split}
    \Gamma^+_{mkln} = &\frac{1}{\hbar^2} \sum_{ij}  \alpha_i\alpha_j\bra{m}Q_{i}\ket{k} \bra{l}Q_{j}\ket{n} \\
    &\times \int_0^{\infty} dt e^{-i\omega_{ln}t}\langle F_{i}(\chi, t)F_{j}(\chi)\rangle \label{eq_Gplus}
\end{split}
\end{equation}
and 
\begin{equation}
\begin{split}
\Gamma^-_{mkln} = &\frac{1}{\hbar^2} \sum_{ij} \alpha_i\alpha_j\bra{m}Q_{j}\ket{k} \bra{l}Q_{i}\ket{n} \\
    &\times  \int_0^{\infty} dt\  e^{-i\omega_{mk}t}\langle F_{j}(\chi)F_{i}(\chi,t)\rangle.\label{eq_Gminus}
    \end{split}
\end{equation}
Here $Q_i$ and $F_i$ are operators from the tunneling Hamiltonian:
\begin{equation}
H_T(\chi) =  \sum_{i=1}^4 \alpha_i Q_iF_{i}(\chi) \label{eq_HT}
\end{equation}
with
\begin{equation}
    F_i({\chi}) = e^{i\sum_{j = L,R}\chi_j N_j}F_i e^{-i\sum_{j = L,R}\chi_j N_j},
\end{equation}
where $N_j$ is the number operator in lead $j$, and
\begin{equation}
\begin{aligned}
\alpha_{1} &= t_{L} & Q_1 &= c^{\dagger} + c & F_{2} &= \psi_{L} \\
\alpha_{2} &= i t_{R} & Q_2 &= c^{\dagger} - c & F_{2} &= \psi_{R} \\
\alpha_{3} &= -t^*_{L} & Q_3 &= c^{\dagger} + c & F_{3} &= \psi^{\dagger}_L \\
\alpha_{4} &= -i t^*_{R} & Q_4 &= c^{\dagger} - c & F_{4} &= \psi^{\dagger}_{R}. \label{eq_HT_def}
\end{aligned}
\end{equation}

\section{Second order tunneling}
Here we consider second order tunneling (SOT) processes starting from the unoccupied state $\ket{0}$ of the wire (the process is similar for the occupied state $\ket{1}$). For low bias voltages the transition $\ket{0} \rightarrow \ket{1}$ is exponentially suppressed with temperature.     We derive an effective low-energy tunneling Hamiltonian, quadratic in the bath operators, effectively taking the SOT processes into account, suppressed only in a power-law fashion in the virtual energy cost.
\subsection{Effective second order tunneling Hamiltonian}
We rewrite our Hamiltonian (Eqs.~(1)--(3) in the main manuscript) \cite{Auerbach1994}
\begin{equation}
\begin{split}
    H &= H_B + H_{\gamma} + H_T \\
    &= H_B + \left(
\begin{array}{c|c}
P_0 (H_{\gamma}+H_T) P_0 & P_0 (H_{\gamma}+H_T) P_1 \\
\hline
P_1 (H_{\gamma}+H_T) P_0 & P_1 (H_{\gamma}+H_T) P_1
\end{array}
\right) 
\end{split}
\end{equation}
where we have introduced the projectors $P_Z = \ket{Z}\bra{Z}$. 
Now we take
\begin{equation}
\begin{split}
    H  &= H_B + \left(
\begin{array}{c|c}
P_0 H_{\gamma} P_0 & P_0 H_T P_1 \\
\hline
P_1 H_T P_0 & P_1 H_{\gamma} P_1 
\end{array}
\right) \\
&= \left(
\begin{array}{c|c}
P_0 (H_B + H_{\gamma}) P_0 & P_0 H_T P_1 \\
\hline
P_1 H_T P_0 & P_1 (H_B + H_{\gamma}) P_1
\end{array}
\right)   
\equiv  \mathcal{H},
\end{split}
\end{equation}
where we wish to approximate the low-energy sector of the matrix.  The resolvent is given by
\begin{equation}
    \mathcal{G}(E) = \frac{1}{E-(H_B + H_{\gamma}+H_T)} = \frac{1}{E-\mathcal{H}}
\end{equation}
and
\begin{equation}
    P_0\mathcal{G}(E)P_0 = P_0(E-\mathcal{H})^{-1}P_0 \equiv (E-H_{eff}^{00})^{-1}
\end{equation}
which defines the new effective Hamiltonian. We find 
\begin{equation}
\begin{split}
    H_{eff}^{00} = &P_0 (H_B+H_{\gamma}) P_0 \\
    & +    P_0 H_T^-\frac{1}{E-(H_B+2\eps)}H_T^+P_0,
    \end{split}
\end{equation}
to second order in $H_T$. We have used 
\begin{equation}
\left[\begin{pmatrix}
    A & B\\
    C& D
\end{pmatrix}^{-1} \right]_{00} = (A-BD^{-1}C)^{-1}
\end{equation}
to calculate $\mathcal{G}(E)$ in the (projected) $\ket{0}\bra{0}$-subspace, and we have introduced
\begin{align}
   \ket{0} \bra{0}H_T^- \ket{1}\bra{1}&= P_0 H_T P_1 \\
   \ket{1}\bra{1} H_T^+ \ket{0}\bra{0} &= P_1 H_T P_0
\end{align}
or, equally,
\begin{align}
     H_T^- &= \sum_{i}(-1)^{i+1}\alpha_i F_i \\
    H_T^+  &= \sum_{i}\alpha_i F_i ,
\end{align}
using the notation from Eq.~(\ref{eq_HT}). The energy $E$ can be written as $E \approx E_{\gamma}+E_B =0+E_0$, where $E_0$ is the bath energy. We then obtain:
\begin{multline}
    H_{eff}^{00}  = P_0 (H_B+H_{\gamma}) P_0 \\
-P_0 H_T^-\frac{1}{H_B - E_0 +2\eps}H_T^+P_0 + \mathcal{O}(H_T^3/\eps^2).
\end{multline}
In total we can now write our full Hamiltonian as
\begin{equation}
\begin{split}
    H & \approx  H_{eff}^{00} + P_1(H_{\gamma}+H_B)P_1 + H_T \\
    &= P_0 (H_B+H_{\gamma}) P_0 -P_0 H_T^-\frac{1}{H_B - E_0+2\eps}H_T^+P_0 \\&+ P_1(H_B + H_{\gamma})P_1 + H_T \\
    &= H_B + H_{\gamma} + H_T -P_0 H_T^-\frac{1}{H_B - E_0+2\eps}H_T^+P_0 \\
    &\equiv H_B + H_{\gamma} + H_T + H_{SOT}^{00},
    \end{split}
\end{equation}
where 
\begin{equation}
    H_{SOT}^{00} = -P_0 \sum_{a,b}(-1)^a\alpha_a \alpha_b F_a \frac{1}{H_B - E_0+2\eps}F_b .
\end{equation}
Repeating the procedure for the $\ket{1}\bra{1}$-sector we arrive at the effective SOT Hamiltonian  $H^{SOT} = H_{SOT}^{00} + H_{SOT}^{11}$ in the main manuscript.

\subsection{Second order tunneling rates}
The SOT rate $W_{00}(\chi)$ can be found from Eq.~(\ref{eq_Wmn}) for $m=n$, but with the $\Gamma^{\pm}_{}$ in Eqs.~(\ref{eq_Gminus})--(\ref{eq_Gplus}) adjusted to account for $H_{SOT}^{00}$ by replacing
\begin{equation}
    \begin{split}
        \alpha_i & \rightarrow (-1)^{a_i} \alpha_{a_i} \alpha_{b_i} \\
        Q_i & \rightarrow -P_0, \ \forall i  \\
        F_i & \rightarrow F_{a_i}  \frac{1}{H_B - E_0+2\eps}F_{b_i},
    \end{split}
\end{equation}
where $a_i,b_i \in \{1,2,3,4\}$.
We write $W_{00}(\chi) = W_{00}^{\chi}(\chi) - W_{00}^{\chi}(\chi = 0) $ and, for simplicity,  present the derivations for $W_{00}^{\chi}$ without the counting fields here and add them later.  To find the new master equation terms, we need to evaluate correlators of four bath operators. 
We find
\begin{widetext}
\begin{multline}
     \Gamma_{0000}^{\pm} = \frac{1}{\hbar}\mathop{\sum_{N_1,N_2}}_{N_3,N_4} \rho^{N_1} \mathop{\sum_{a_i,a_j}}_{b_i,b_j} \alpha_{a_i}\alpha_{b_i}\alpha_{a_j}^*\alpha_{b_j}^* (-1)^{a_i+a_j} \\
     \times \int dt\ e^{\pm i\frac{E_{N_1}-E_{N_3}}{\hbar}t} \frac{\bra{N_1}F_{a_i}\ket{N_2} \bra{N_2}F_{b_i}\ket{N_3}   }{H_B^{N_1}-H_B^{N_2}-2\eps} \frac{\bra{N_3}F_{a_j}\ket{N_4} \bra{N_4}F_{b_j}\ket{N_1}  } {H_B^{N_1}-H_B^{N_4}-2\eps} \\
=\frac{1}{\hbar}\sum_{N_1,N_3} \rho^{N_1}\int dt \ e^{\pm i\frac{E_{N_1}-E_{N_3}}{\hbar}t} \left(\sum_{N_2} \sum_{a_i,b_i} \alpha_{a_i} \alpha_{b_i} (-1)^{a_i} \frac{\bra{N_1}F_{a_i}\ket{N_2} \bra{N_2}F_{b_i}\ket{N_3}    }{H_B^{N_1}-H_B^{N_2}-2\eps}  \right) \\
 \times \left( \sum_{N_4}\sum_{a_j, b_j} \alpha_{a_j}^* \alpha_{b_j}^* (-1)^{a_j}\frac{\bra{N_3}F_{a_j}\ket{N_4} \bra{N_4}F_{b_j}\ket{N_1}  } {H_B^{N_1}-H_B^{N_4}-2\eps} \right)
     \\
     =\frac{1}{\hbar}\sum_{N,M} \rho^N\delta(E_N-E_M) \int dt\  e^{\pm i\frac{E_{N_1}-E_{N_3}}{\hbar}t}  \left| \mathop{\sum_{a_i, b_i}}_{N_2} \alpha_{a_i}\alpha_{b_i} (-1)^{a_i} \frac{\bra{N}F_{a_i}\ket{N_2}\bra{N_2}F_{b_i}\ket{M}}{E_N - E_{N_2}-2\eps}  \right|^2.
\end{multline}
Using $W_{00}^{\chi} = \Gamma_{0000}^+ + \Gamma_{0000}^-$, we can perform the time integral to obtain
\begin{equation}
    W_{00}^{\chi}(0) = \frac{2\pi}{\hbar}\sum_{N,M} \rho^{N} \delta (E_{N}-E_{M})  \left| \mathop{\sum_{a_i, b_i}}_{N_2} \alpha_{a_i}\alpha_{b_i} (-1)^{a_i} \frac{\bra{N}F_{a_i}\ket{N_2}\bra{N_2}F_{b_i}\ket{M}}{E_N - E_{N_2}-2\eps}  \right|^2.
\end{equation}
Note that this has the form of a Fermi golden rule rate in second-order perturbation theory.
\end{widetext}
We do not consider local Andreev reflection processes in the following, because they are subleading contributions to the crossed Andreev reflection processes in the regime we discuss in the main text, as is also known from previous work \cite{Nilsson2008}.
Using
\begin{equation}
\psi_{L,R} = \frac{1}{\sqrt{L}}\sum_k e^{ikx} c_{L,R; k},
\end{equation}
where $L$ is a quantization length of the leads, we can write down four possibilities for the non-local processes:
\begin{equation}
\begin{split}
\ket{N} &= \cdag_{L;k}\cdag_{R,k'}\ket{M} \\
&\rightarrow \rho^N = f_L(E_k)f_R(E_{k'})\\
\ket{N} &= \cdag_{L;k}c_{R,k'}\ket{M} 
    \\ & \rightarrow \rho^N = f_L(E_k)(1-f_R(E_{k'}))\\
\ket{N} &= c_{L;k}\cdag_{R,k'}\ket{M}
    \\ & \rightarrow \rho^N  = (1-f_L(E_k))f_R(E_{k'})\\
\ket{N} &= c_{L;k}c_{R,k'}\ket{M}
    \\ &\rightarrow \rho^N  = (1-f_L(E_k))(1-f_R(E_{k'}))
\end{split}
\end{equation}

(where each process can happen in two distinct ways). We  write $W_{00}^{\chi}(0)  = W_{00}^{++} + W_{00}^{+-} + W_{00}^{-+} + W_{00}^{--}$ where
\begin{multline}
W_{00}^{++} = \frac{2\pi}{\hbar} \frac{1}{L}\rho_R|t_L|^2|t_R|^2 \sum_k f_L(E_k)f_R(-E_{k})\\
     \times \left| \frac{1}{E_k -2\eps}- \frac{1}{E_k +2\eps}  \right|^2,
     \end{multline}
     \begin{multline}
W_{00}^{+-} = \frac{2\pi}{\hbar}\frac{1}{L}\rho_R |t_L|^2|t_R|^2 \sum_k f_L(E_k)(1-f_R(E_{k}))\\
  \times \left| \frac{1}{E_k -2\eps}- \frac{1}{E_k +2\eps}  \right|^2,
       \end{multline}
     \begin{multline}
W_{00}^{-+} = \frac{2\pi}{\hbar}\frac{1}{L}\rho_R |t_L|^2|t_R|^2 \sum_k (1-f_L(E_k))f_R(E_{k}) \\
   \times \left| \frac{1}{E_k -2\eps}- \frac{1}{E_k +2\eps}  \right|^2
        \end{multline}
and
     \begin{multline}
W_{00}^{--} = \frac{2\pi}{\hbar}\frac{1}{L}\rho_R |t_L|^2|t_R|^2 \sum_k (1-f_L(E_k))(1-f_R(-E_{k})) \\
 \times \left| \frac{1}{E_k -2\eps}- \frac{1}{E_k +2\eps}  \right|^2.\\
\end{multline}
(where $W_{00}^{+-} = W_{00}^{-+}$ for a symmetric bias).
Now we take the sum over $k$ to an integral over energy $\eps_L$, insert $+i\gamma$ in the denominator and expand the absolute value squared as
\begin{multline}
   \left| \frac{1}{E_k -2\eps}- \frac{1}{E_k +2\eps}  \right|^2 \rightarrow \left| \frac{1}{\eps_L -2\eps + i\gamma}- \frac{1}{\eps_L +2\eps + i\gamma}  \right|^2\\
=  \left[ \frac{1}{(\eps_L + 2\eps)^2 + \gamma ^2}  + \frac{1}{(\eps_L - 2\eps)^2 + \gamma ^2} \right. \\
\left.- 2\mathrm{Re}\left(\frac{1}{\eps_L- 2\eps + i\gamma}\frac{1}{\eps_L+ 2\eps - i\gamma}\right) \right] 
\end{multline}
Finally, the SOT rate 
can be written as 
\begin{multline}
\label{SOT_rate}
    W_{00}^{\chi}(0) = \frac{2\pi}{\hbar}  |t_L|^2|t_R|^2 \rho_R\rho_L \\
    \times\int d\eps_L\left[f_L(\eps_L)f_R(-\eps_L) + f_L(\eps_L)(1-f_R(\eps_L))\right. \\
    \left.+ (1-f_L(\eps_L))f_R(\eps_L) + (1-f_L(\eps_L)(1-f_R(-\eps_L)) \right]\\
\times\left[ \frac{1}{(\eps_L + 2\eps)^2 + \gamma ^2}  + \frac{1}{(\eps_L - 2\eps)^2 + \gamma ^2} \right. \\
\left.- 2\mathrm{Re}\left(\frac{1}{\eps_L- 2\eps + i\gamma}\frac{1}{\eps_L+ 2\eps - i\gamma}\right) \right]
    .
\end{multline}

\subsection{Regularization and analytical solutions of the integrals}


To solve the integrals for the SOT rates, we employ a regularization scheme, physically corresponding to introducing a finite width/lifetime $\gamma$ that the tunneling electron experiences. The integration is carried out over the whole real axis, and we convert it to an integral over the complex plane and use Cauchy's theorem to instead calculate the residue of the poles.
For the integrals of the general forms
\begin{equation}
    \int d\eps_L \frac{F(\eps_L)}{(\eps_L\pm 2\eps)^2+ \gamma^2},
\end{equation}
(where $F(\eps_L) \in \{ \alpha, \beta, \delta\}$ is a Fermi function expression), we follow Ref.~\cite{Koch2004,Koch2006} in the approach of removing the lowest order $1/\gamma$. Note that this is a crucial step for the integrals to converge.

The integral can then be performed
\begin{equation}
    \int d\eps_L \frac{F(\eps_L)}{(\eps_L\pm 2\eps)^2+ \gamma^2}
\equiv  2\pi i\frac{i}{2\gamma} f(\gamma),
\end{equation}
where $f(\gamma)$ is some function obtained from performing the integral.
We expand $f(\gamma)$ and subtract the leading order term that is $\mathcal{O}(1/\gamma)$. Then we take $\gamma \rightarrow 0$, yielding:
\begin{equation}
\begin{split}
    & 2\pi i\frac{i}{2\gamma} f(\gamma) -\mathcal{O}(1/\gamma) \\
    = &2\pi i\frac{i}{2\gamma} \sum_{n=0}^{\infty} \frac{f^{(n)}(\gamma)}{n!}\gamma^n -\mathcal{O}(1/\gamma) \\
    = &2\pi i\frac{i}{2\gamma} \sum_{n=1}^{\infty} \frac{f^{(n)}}{n!}\gamma^n = 2\pi i\frac{i}{2} f'(\gamma= 0)
    \end{split}
\end{equation}
where $\gamma = 0$ for the last equality.

For a symmetric bias $\mu_L = \mu_R = \mu $ we find the following expressions for the integrals (using results from Refs.~\cite{Koch2004,Koch2006}):
\begin{widetext}
\begin{equation}
    \begin{split}
& \int d\eps_L \frac{f_L(\eps_L)f_R(-\eps_L)}{(\eps_L + 2\eps)^2 + \gamma ^2} = 
\frac{\beta}{2} \frac{\pi}{2} n_B(-2\mu)\left[\sech ^2(\beta\frac{\mu-2\eps}{2})- \sech ^2(\beta\frac{\mu+2\eps}{2})\right]   \\
& \int d\eps_L \frac{ f_L(\eps_L)(1-f_R(\eps_L))}{(\eps_L + 2\eps)^2 + \gamma ^2} = 
 -2\pi\frac{\beta }{8}\sech^2(\beta\frac{\mu+2\eps}{2})\tanh(\beta\frac{\mu+2\eps}{2})\\
& \int d\eps_L \frac{(1-f_L(\eps_L))(1-f_R(-\eps_L) }{(\eps_L + 2\eps)^2 + \gamma ^2}  = 
\frac{\beta}{2} \frac{\pi}{2} n_B(2\mu)\left[\sech ^2(\beta\frac{\mu+2\eps}{2})- \sech ^2(\beta\frac{\mu-2\eps}{2})\right] \\
&  \int d\eps_L \frac{f_L(\eps_L)f_R(-\eps_L)}{(\eps_L - 2\eps)^2 + \gamma ^2} = 
   \frac{\beta}{2} \frac{\pi}{2} n_B(-2\mu)\left[\sech ^2(\beta\frac{\mu+2\eps}{2})- \sech ^2(\beta\frac{\mu-2\eps}{2})\right]       \\
& \int d\eps_L \frac{ f_L(\eps_L)(1-f_R(\eps_L))}{(\eps_L - 2\eps)^2 + \gamma ^2}  = 
 -2\pi\frac{\beta }{8}\sech^2(\beta\frac{\mu-2\eps}{2})\tanh(\beta\frac{\mu-2\eps}{2})\\
& \int d\eps_L \frac{(1-f_L(\eps_L))(1-f_R(-\eps_L)}{(\eps_L - 2\eps)^2 + \gamma ^2}  = 
\frac{\beta}{2} \frac{\pi}{2} n_B(2\mu)\left[\sech ^2(\beta\frac{\mu-2\eps}{2})- \sech ^2(\beta\frac{\mu+2\eps}{2})\right] \\
& \mathrm{Re}\int d\eps_L \frac{f_L(\eps_L)f_R(-\eps_L)}{(\eps_L- 2\eps + i\gamma)(\eps_L+ 2\eps - i\gamma)} = 
\frac{1}{2\eps} n_B(-2\mu)\mathrm{Re}\left[ \psi (\frac{1}{2}- i\beta\frac{\mu+2\eps}{2\pi}) -\psi (\frac{1}{2}+i\beta\frac{\mu-2\eps}{2\pi})   \right]\\
&  \mathrm{Re}\int d\eps_L \frac{ f_L(\eps_L)(1-f_R(\eps_L))}{(\eps_L- 2\eps + i\gamma)(\eps_L+ 2\eps - i\gamma)} = \frac{1}{4\eps} \frac{\pi}{4} \left[\sech ^2(\beta\frac{\mu-2\eps}{2})+ \sech ^2(\beta\frac{\mu+2\eps}{2})\right]
 \\
& \mathrm{Re}\int d\eps_L \frac{ (1-f_L(\eps_L))(1-f_R(-\eps_L)}{(\eps_L- 2\eps + i\gamma)(\eps_L+ 2\eps - i\gamma)} = \frac{1}{2\eps} n_B(2\mu)\mathrm{Re}\left[ \psi (\frac{1}{2}+ i\beta\frac{\mu-2\eps}{2\pi}) -\psi (\frac{1}{2}-i\beta\frac{\mu+2\eps}{2\pi})   \right],
    \end{split}
    \label{eq_ints}
\end{equation}
\end{widetext}
where we have used the Bose distribution function
\begin{equation}
    n_B(x) = \frac{1}{e^{x\beta}-1}
\end{equation}
with $\beta = \frac{1}{kT}$
and $\psi (z) $ is the digamma function. 
In total, including counting fields, we find
\begin{multline}
    W_{00}^{\chi}(\chi)   = 
     W_{00}^{++} e^{i(\chi_L+\chi_R)}  + W_{00}^{+-}e^{i(\chi_L-\chi_R)} \\
     + W_{00}^{-+} e^{-i(\chi_L-\chi_R)}+  W_{00}^{--} e^{-i(\chi_L+\chi_R)} .  
    \end{multline}
\begin{figure}
 \vspace{0.3cm}
  \begin{flushleft}
\hspace{0.5cm}\large{(a)} \hspace{3.3cm} \large{(b)}
\end{flushleft} 
 \vspace{-0.3cm}
 \centering
 \includegraphics[width=0.2495\textwidth, trim=2.1cm 6.6cm 3.25cm 6cm, clip]{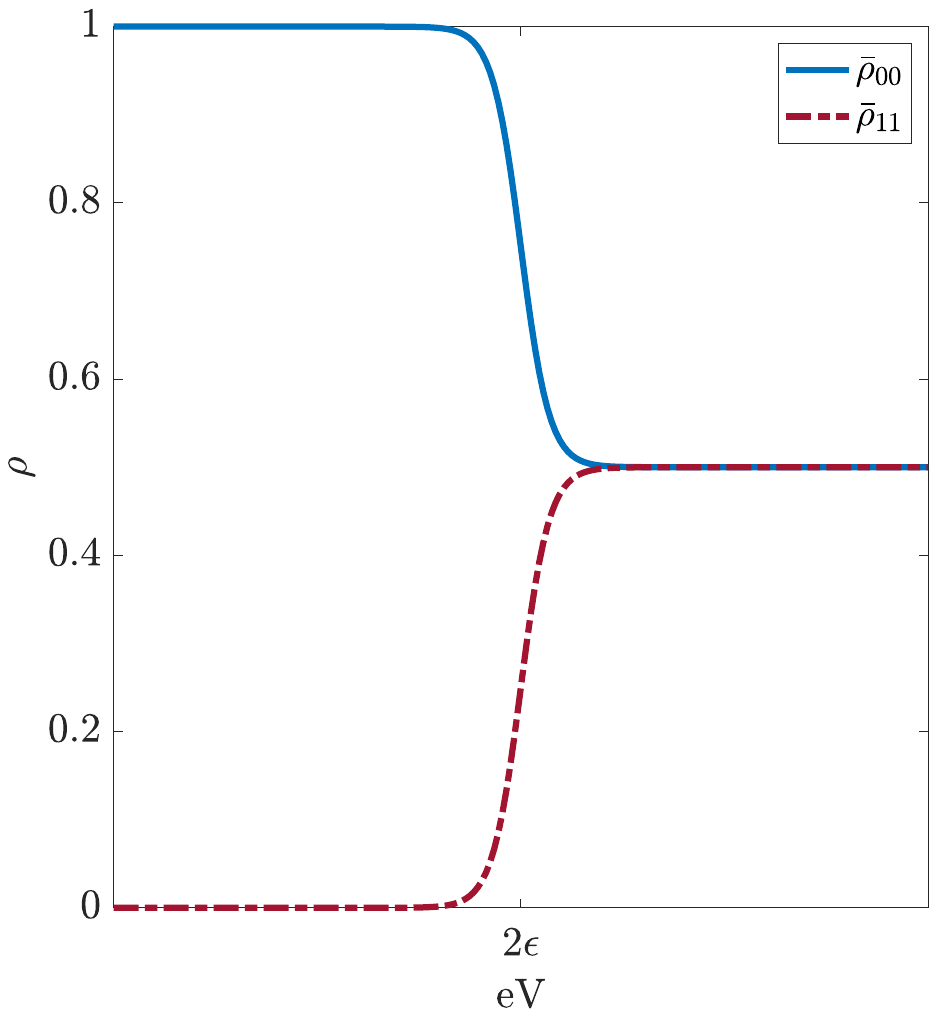} \includegraphics[width=0.223\textwidth, trim=3.7cm 6.6cm 3.25cm 6cm, clip]{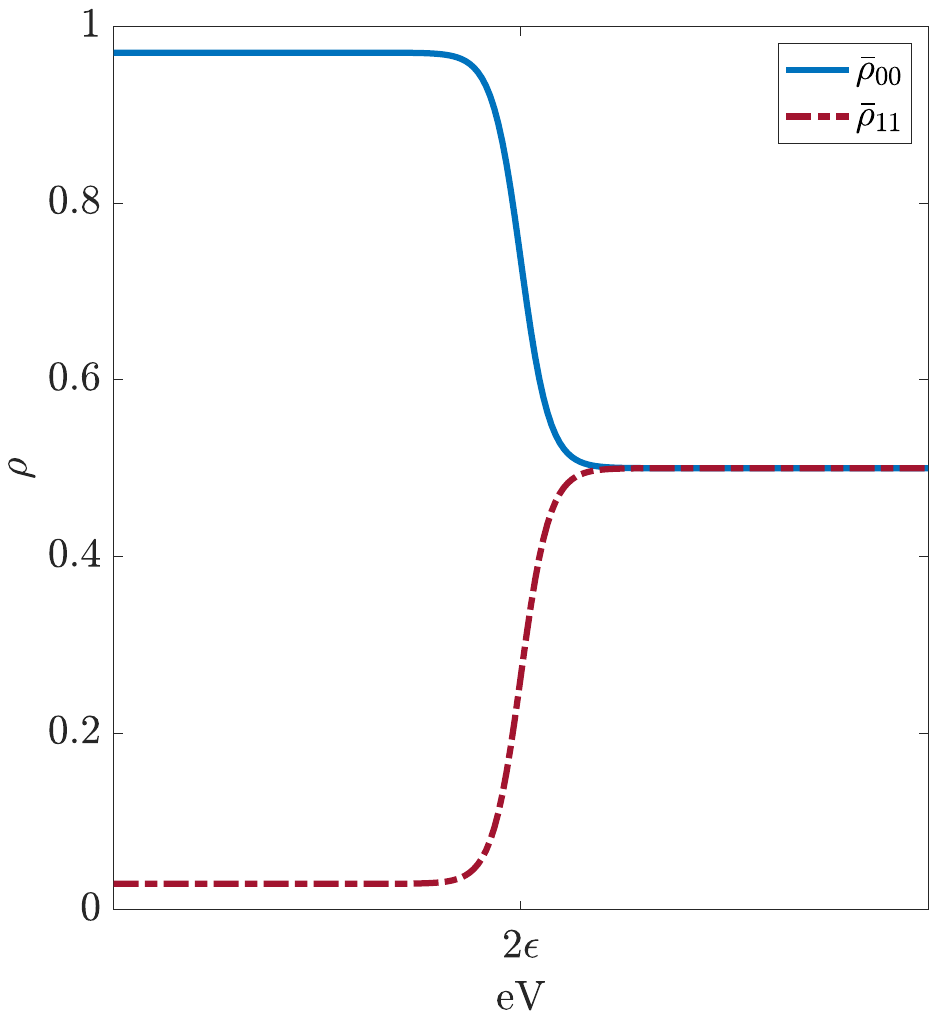}
    \caption{Steady state solutions for the Master equation (a) without QPP and (b) with QPP. Parameters are: $k_BT = 0.07\eps$, $\Gamma_i=2.5\cdot 10^{-2}\eps$, ($i \in \{L,R\}$) and $\Gamma_{QP}=1\cdot 10^{-3}\eps$.}
    \label{fig_ss}
\end{figure}
We can find the steady state solution $\bar{\rho} = (\bar{\rho}_{00} \ \bar{\rho}_{11})^T$ to the Master equation $\mathcal{L}\bar{\rho}=0$. The steady state solution is shown in Fig.~\ref{fig_ss} for the cases with zero and finite $\Gamma_{QP}$. We see that a finite QPP rate yields a non-zero probability for the system to be in excited state $\ket{1}$, in contrast to the case when $\Gamma_{QP}=0$, where population of $\ket{1}$ is exponentially suppressed with temperature in the regime of interest ($k_BT,\mu \ll 2\epsilon$).

We note that the integral in Eq.~(\ref{SOT_rate}) for $W_{00}^{\chi}(0)$ can be conveniently evaluated at $k_BT=0$ where only the term containing $f_L(\epsilon_L)f_R(-\epsilon_L)=\theta(\mu-\epsilon_L)\theta(\mu+\epsilon_L)$ contributes describing CAR-processes. Here, $\theta(x)$ is the Heaviside step-function. To leading order in $\mu/\epsilon$, we can neglect $\epsilon_L$ compared to $2\epsilon$ in the integrand and obtain
\begin{equation}
\label{zerotemperatureCAR}
    W_{00}^{\chi}(0)=W_{00}^{++}=\frac{1}{\pi\hbar}\frac{\Gamma_{L}\Gamma_{R}}{\epsilon^2}\mu,\quad k_BT=0.
\end{equation}
As we explicitly show in Eq.~(\ref{STcurrent}) of this supplemental material, ST rates do not contribute an average current $I_j$ at zero temperature and $I_j=e({\bar \rho}_{00} W_{00}^{\chi}(0)+{\bar \rho}_{11} W_{11}^{\chi}(0))$. We remark that $W_{11}^{\chi}(0)$ is the same as $W_{00}^{\chi}(0)$ with $\epsilon\rightarrow -\epsilon$, and therefore $W_{11}^{\chi}(0)$ agrees with Eq.~(\ref{zerotemperatureCAR}) to leading order in $\mu/\epsilon$ and $k_BT=0$. Therefore, the average current at $k_BT=0$ and to leading order in $\mu/\epsilon$ becomes 
\begin{equation}
\label{zerotemperatureCARcurrent}
    I_j=\frac{e}{\pi\hbar}\frac{\Gamma_{L}\Gamma_{R}}{\epsilon^2}\mu, \quad k_BT=0,
\end{equation}
as ${\bar \rho}_{00}+{\bar \rho}_{11}=1$. This result agrees with the elastic scattering matrix calculation~\cite{Nilsson2008}, and, as we show below, is independent of $\Gamma_{QP}$ in the limit considered here.
 
\section{Transport properties}

\subsection{Full counting statistics: counting fields}

To find the cumulant generating function (CGF) $S(\chi,t)$, we use the long-time behaviour \cite{Probst2022} 
    \begin{equation}
        S(\chi,t) \approx \Lambda_0 (\chi) t,
    \end{equation}
where $\Lambda_0 $ is the eigenvalue of $\mathcal{L}$ with the smallest real part (the eigenvalue that is continuously connected to zero for $\chi=0$). This eigenvalue can be found either by solving the characteristic equation for $\mathcal{L}$ or
by the use of Rayleigh-Schr\"odinger perturbation theory \cite{Probst2022,Flindt2010}.

The total Liouvillian, including ST and SOT, as well as QPP, can be written in matrix form as
\begin{equation}
\label{Leq}
\begin{split}
    \mathcal{L}_{tot} =& \mathcal{L}_{ST} + \mathcal{L}_{SOT} + \mathcal{L}_{QP} \\
    =&
     \begin{pmatrix}
   \gamma_{00} - W_{00} + \Gamma_{QP}  & -W_{01} - \Gamma_{QP} \\
     -W_{10} - \Gamma_{QP} & \gamma_{11} - W_{11} + \Gamma_{QP}
    \end{pmatrix},
    \end{split}
\end{equation}
where we have introduced $\gamma_{00} = W_{10}(\chi = 0)$ and $\gamma_{11} = W_{01}(\chi = 0)$. 
\begin{widetext}
We find that the eigenvalue connected to zero is given by
\begin{equation}
\begin{split}
       \Lambda_0 =& \frac{\gamma_{00}+ \gamma_{11}-W_{00}-W_{11}+2\Gamma_{QP} }{2}  - \frac{1}{2}\sqrt{(\gamma_{00}-\gamma_{11} - W_{00}+W_{11})^2+4(W_{10}+\Gamma_{QP} )(W_{01}+\Gamma_{QP})}.
\end{split}
\end{equation}
The current in the left (right) lead is proportional to the derivative of $\Lambda_0$ with respect to $\chi_L$ ($\chi_R$), taken at $\chi = 0$:
\begin{equation}
\label{currentexact}
     \frac{\partial \Lambda_0}{\partial \chi_L} =  -\frac{1}{2}\frac{\partial (W_{00}+W_{11})}{\partial\chi_L} - 
     \frac{\frac{\partial (W_{11}-W_{00})}{\partial\chi_L}(\gamma_{11}-\gamma_{00}- W_{00}+W_{11}) + 2\frac{\partial W_{10}}{\partial\chi_L}(W_{01}+\Gamma_{QP}) + 2 (W_{10}+\Gamma_{QP})\frac{\partial W_{01}}{\partial\chi_L}}{2\sqrt{(\gamma_{00}-\gamma_{11}- W_{00}+W_{11})^2+4(W_{10}+\Gamma_{QP} )(W_{01}+\Gamma_{QP})} } 
\end{equation}
The local noise is proportional to the second derivative with respect to the same field:
\begin{equation}
\label{localexakt}
     \frac{\partial ^2\Lambda_0}{\partial\chi_L^2}  = -\frac{1}{2}\frac{\partial^2 (W_{00}+W_{11})}{\partial\chi_L^2} - \frac{1}{2B}\left[ \frac{\partial b_L}{\partial\chi_L}-\frac{{b_L}^2}{{B}^2}\right],
\end{equation}
where we have introduced
    \begin{equation}
    \label{Bparameter}
    B = \sqrt{(\gamma_{00}-\gamma_{11} - W_{00}+W_{11})^2+4(W_{10}+\Gamma_{QP} )(W_{01}+\Gamma_{QP})} ,
\end{equation}
\end{widetext}
and 
\begin{multline}
\label{blparameter}
      b_L = \frac{\partial (W_{11}-W_{00})}{\partial\chi_L}(\gamma_{11}-\gamma_{00}- W_{00}+W_{11}) \\
     + 2\frac{\partial W_{10}}{\partial\chi_L}(W_{01}+\Gamma_{QP})  + 2 (W_{10}+\Gamma_{QP})\frac{\partial W_{01}}{\partial\chi_L}.  
\end{multline}
The cross-noise can be found from the mixed second derivative 
\begin{equation}     
     \frac{\partial ^2\Lambda_0}{\partial\chi_L\partial\chi_R}  = -\frac{1}{2} \frac{\partial ^2(W_{00}+W_{11})}{\partial\chi_L\partial\chi_R}  -\frac{1}{2} \frac{\partial^2 B}{\partial\chi_L\partial \chi_R}. 
\end{equation}

\section{Influence of non-local tunneling amplitudes}

In this section, we discuss the influence of non-local tunnel couplings on the transport properties. We introduce a parameter $\phi$ in the tunneling Hamiltonian $H_T$ that interpolates between the coupling to a single Majorana bound state (MBS, $\phi=0$) and the coupling to a single fermion ($\phi=\pi/4$). We show that for any finite $\phi$, there exists a threshold bias voltage, below which the local shot noise becomes qualitatively different from the single-MBS coupling case considered in the main text. With a generic $\phi$, the coupling to a usual Andreev bound state can be modeled.

\subsection{Tunneling Hamiltonian and tunneling rates with non-local couplings}
The Hamiltonian describing tunneling between the leads and the MBSs including non-local couplings can be written as \cite{Schuray2017}
\begin{equation}
\label{se0}
H_T=i\sum_{j=L,R}t_j\psi(\cos{\phi}\gamma_1+i\sin{\phi}\gamma_2)+h.c.
\end{equation}
The case $\phi=0$ corresponds to the case of coupling to a single MBS considered in the main text. Note that the case $\phi=\pi/4$ leaves terms of the form $it_{j}\psi c +h.c.$ which represents the tunneling of a lead electron to a Dirac fermion (being not a superposition of a creation and annihilation operator) and leads to no transport without QPP. For $|\phi|$ between 0 and $\pi/4$, Eq.~(\ref{se0}) constitutes a minimal model for the coupling to an ordinary Andreev bound state. 

Leaving $\phi$ as a parameter, we can calculate the counting field-dependent first-order tunneling rates, assuming for simplicity equal tunneling rates for the two leads $\Gamma_L=\Gamma_R=\Gamma$, as (we set $\hbar=1$)
\begin{equation}
\label{se1}
 W_{01}=(1+\sin(2\phi))W_{01}^{+}+(1-\sin(2\phi))W_{01}^{-},
\end{equation}
where $W_{01}^{+}=\Gamma(1-f(\Delta_{01}))(e^{-i\chi_L}+e^{-i\chi_R})$, $W_{01}^{-}=\Gamma f(-\Delta_{01})(e^{i\chi_L}+e^{i\chi_R})$, and
\begin{equation}
\label{se2}
 W_{10}=(1-\sin(2\phi))W_{10}^{+}+(1+\sin(2\phi))W_{10}^{-},
\end{equation}
where $W_{10}^{+}=\Gamma(1-f(\Delta_{10}))(e^{-i\chi_L}+e^{-i\chi_R})$, $W_{10}^{-}=\Gamma f(-\Delta_{10})(e^{i\chi_L}+e^{i\chi_R})$. Here, $\Delta_{01}=-\Delta_{10}=E_1-E_0$ is the energy difference between the two system states and $W_{mn}^{\eta}$ are the tunneling rates going from state $n$ to state $m$ thereby adding ($\eta=+$) or subtracting ($\eta=-$) a particle in one of the leads.

As in the main text, we are interested in the low energy regime $k_BT,eV \ll \Delta_{01}$, where we can replace $(1-f(\Delta_{01}))\approx f(\Delta_{10})\approx 1$. In this limit, $W_{10}\approx 0$. Further,
Eqs.~(\ref{se1}) and (\ref{se2}) show that for a general $\phi$, the rate to add (subtract) a particle to (from) a lead is not the same anymore (except for the MBS coupling case $\phi=0$) and independent on the bias voltage $V$. Therefore, the average current and noise will be dominated by first order rates $W_{01}^{\pm}$ for $\phi\neq 0$ if $\Gamma_{QP}$ is finite. We therefore can neglect SOT-rates for small coupling $\Gamma$ and/or small enough $V$. 
\subsection{Transport properties with non-local couplings}
Using only ST-rates in the Liouvillian Eq.~(\ref{Leq}), the average current $I_j$ in lead $j=L,R$ becomes $I_j=-ie((\partial W_{10}/\partial\chi_j) {\bar \rho_{00}}+(\partial W_{01}/\partial\chi_j) {\bar \rho_{11})}\big|_{\chi=0}$ with the steady state solutions for the occupation probabilities of the two system states $|0\rangle$ and $|1\rangle$
\begin{equation}
\label{rhog}
{\bar \rho_{00}}=\frac{W_{01}+\Gamma_{QP}}{W_{01}+W_{10}+2\Gamma_{QP}}\big|_{\chi=0},
\end{equation}
\begin{equation}
\label{rhoe}
{\bar \rho_{11}}=\frac{W_{10}+\Gamma_{QP}}{W_{01}+W_{10}+2\Gamma_{QP}}\big|_{\chi=0}.
\end{equation}
For the low-energy regime with the approximations given above, we obtain (${\bar \rho_{11}}|_{\chi=0}=\Gamma_{QP}/(4\Gamma+2\Gamma_{QP})$)
\begin{equation}
\label{STcurrent}
\begin{split}
    I_{j}(V\rightarrow 0)&=-ie(\partial W_{01}/\partial\chi_j) {\bar \rho_{11}}\big|_{\chi= 0}\\
    &=2e\Gamma\sin(2|\phi|){\bar \rho_{11}}\big|_{\chi=0}.
\end{split}
\end{equation}
This result shows that the ST average current is finite at zero voltage only if QPP and non-local tunnel couplings (i.e., for $\phi\neq 0$) are present. 

For the local noise the low-energy result for $V\rightarrow 0$ becomes
\begin{equation}
\label{localnoisephi}
\begin{split}
 S_{jj}(V\rightarrow 0)&=-e^2\frac{\partial^{2}W_{01}}{\partial \chi_j^2}{\bar \rho}_{11}\big|_{\chi=0}+2e^2\frac{(\frac{\partial W_{01}}{\partial \chi_{j}})^2{\bar \rho_{11}}^2}{W_{01}+2\Gamma_{QP}}\big|_{\chi=0}\\
 &=2e^2\Gamma{\bar \rho_{11}}\left(1-\frac{2\Gamma\sin^2(2\phi)}{2\Gamma+\Gamma_{QP}}{\bar \rho_{11}}\right)\big|_{\chi=0}.\\
\end{split}
\end{equation}
Note that the noise is the sum of two terms. The first term is simply the shot noise generated by tunneling over the barriers (as an electron or a hole) and is independent of $\phi$, whereas the second term accounts for correlations of subsequent tunneling events \cite{Probst2022}, which, on average, are finite if $\phi\neq 0$. Therefore, the local Fano factor $F_{jj}=S_{jj}/(eI_{j})$ becomes 
\begin{equation}
F_{jj}=\frac{1-\frac{2\Gamma\sin^2(2\phi)}{2\Gamma+\Gamma_{QP}}{\bar \rho_{11}}}{\sin(2|\phi|)}\big|_{\chi=0}
\approx\frac{1}{\sin(2|\phi|)}.
\end{equation}
The approximate result applies for weak QPP ($\Gamma_{QP}\ll \Gamma$).
Note that this very appealing result applies only when the SOT events (CAR process for small voltages) are negligible, i.e., for small $\Gamma$ and/or small bias voltages $V$. 
More precisely, for any $\phi$ there exists a threshold voltage $V_T$ (depending also on other parameters) below which the local Fano factor $F_{jj}\approx 1/\sin(2|\phi|)$ is constant. For $\phi=0$ (the case of coupling to a single MBS), the local Fano factor $F_{jj}\approx 1/V$ as $V\rightarrow 0$ and $V_T=0$ (here we neglect contributions coming from thermal noise).

In conclusion, we predict that the existence of nonlocal tunneling amplitudes simulating the case of coupling to an ordinary Andreev bound state lead to a threshold bias voltage $V_T$ below which the local Fano factor $F_{jj}$ becomes constant as a function of $V$ in the presence of finite QPP. On the contrary, as considered in the main text, $F_{jj}$ shows a power-law in $V$ down to arbitrarily small bias voltages (i.e., $V_T=0$) when $H_T$ couples to a single MBSs ($\phi=0$). The peculiar power-law voltage behavior in $F_{jj}$ observed in the main text is therefore unique to the case of coupling to a single MBS in the presence of QPP.

\section{Low-energy form of the local Fano factor $F_{LL}$}
In this section, we derive the low-energy form of the local Fano factor $F_{LL}$ quoted on page 4 of the main text.
\subsection{Local Fano factor $F_{LL}$ at small temperatures}
For $k_BT\ll\mu\ll2\epsilon$ with $\mu_L=\mu_R=\mu$, we can obtain a low-energy form for $F_{jj}$ ($j=R,L$) as follows (we consider $\phi=0$ and set $\hbar=1$ as in the main text).

The average current $I_j$ is governed by CAR-processes, since the ST-current
vanishes (cf. Eq.~(\ref{STcurrent})). The result is $I_j=(e/\pi)\mu\Gamma_{L}\Gamma_{R}/\epsilon^2$ (cf. Eq.~(\ref{zerotemperatureCARcurrent})). 

If the system is in state $|0\rangle$ with probability ${\bar \rho}_{00}$ the noise is governed by independent CAR-processes involving the SOT-rate $W_{00}$ with associated local noise $eI_j$. However, if the system is in state $|1\rangle$ with probability ${\bar \rho}_{11}$ the noise $S_{jj}$ originates from two channels, independent CAR-processes involving now the SOT-rate $W_{11}$ with the same noise $eI_j$ and the much stronger contribution from the ST-rate $W_{01}$ for small tunneling amplitudes (i.e., for $\Gamma^2\mu/\epsilon^2\ll \Gamma$) which adds the noise contribution $\approx 2e^2\Gamma_j$ (cf. Eq.~(\ref{localnoisephi})). 

For the local Fano factor $F_{jj}=S_{jj}/(eI_{j})$ we therefore obtain $F_{jj}\approx ({\bar \rho}_{00}eI_j+{\bar \rho}_{11}(eI_j+2\Gamma_{j}e^2)/(eI_j)$, and explicitly
\begin{equation}
\label{Fanolocal}
    F_{jj}\approx 1+{\bar \rho}_{11}\frac{2\pi\epsilon^2\Gamma_{j}}{\Gamma_{L}\Gamma_{R}\mu},
\end{equation}
since ${\bar \rho}_{00}+{\bar \rho}_{11}=1$.
In the regime $k_BT,\mu\ll 2\epsilon$, Eq.~(\ref{rhoe}) becomes independent of $\mu$ and $k_BT$, i.e.,
\begin{equation}
\label{r1}
   {\bar \rho}_{11}=\frac{\Gamma_{QP}}{2(\Gamma_{L}+\Gamma_{R})+2\Gamma_{QP}}.
\end{equation}
We note that we can derive the result Eq.~(\ref{Fanolocal}) from Eqs.~(\ref{currentexact})-(\ref{blparameter}) which becomes exact in the stated limit above (i.e., for $k_BT\ll\mu\ll2\epsilon$).

\subsection{Limit of strong poisoning}
Let us consider the limit of strong QPP and/or weak tunnel-coupling $\Gamma\ll\Gamma_{QP}$. For simplicity, we set $\Gamma_L=\Gamma_R=\Gamma$. Using Eqs.~(\ref{Fanolocal}) and (\ref{r1}) and $\Gamma\ll\Gamma_{QP}$, we obtain
\begin{equation}
\label{Fanolocal2}
    F_{jj}\approx 1+\frac{\pi\epsilon^2}{\Gamma\mu}\approx \frac{\pi\epsilon^2}{\Gamma\mu},
\end{equation}
which is much larger than 1 as $\Gamma\mu\ll\epsilon^2$. The maximal possible $F_{jj}$ can be estimated by setting $\mu\simeq k_BT$ with $F_{jj}\approx\pi\epsilon^2/(\Gamma k_BT)$. 
\bibliography{MFs_QPP_paper.bib}